\documentclass{ws-ijmpa}
\usepackage{url}
\usepackage[super,compress]{cite}
\usepackage{graphicx, epsfig, amssymb} 
\usepackage{amsmath, amsfonts}
\usepackage{bm} 
\usepackage{color}
\usepackage[dvips,breaklinks]{hyperref}
\hypersetup{colorlinks,urlcolor=black,citecolor=black,linkcolor=black,filecolor=black}
\usepackage{breakurl}

\def\nn{\nonumber}

\def\th{\vartheta}
\def\cQ{{\cal Q}}
\def\lp{{\ell+1}}
\def\lm{{\ell-1}}
\def\lpp{{\ell+2}}
\def\lmm{{\ell-2}}

\begin{document}

\markboth{Paolo Pani}
{Advanced Methods in Black-Hole Perturbation Theory}

%
\catchline{}{}{}{}{}
%

\title{Advanced Methods in Black-Hole Perturbation Theory\footnote{Based on a series of lectures given at the NR/HEP2: Spring School [11-14 March, 2013 (Lisbon, Portugal)]. {\scshape Mathematica}\textsuperscript{\textregistered} notebooks are publicly available on the School webpage~\cite{webpage}.}}

\author{Paolo Pani}

\address{CENTRA, Departamento de F\'{\i}sica, 
Instituto Superior T\'ecnico,\\ Universidade T\'ecnica de Lisboa - UTL,
Avenida~Rovisco Pais 1, 1049 Lisboa, Portugal\\
\&\\
Institute for Theory $\&$ Computation, Harvard-Smithsonian
CfA, 60 Garden Street, Cambridge, MA, USA
\\\vspace{0.5cm}
paolo.pani@ist.utl.pt}

\maketitle

%

\begin{abstract}
Black-hole perturbation theory is a useful tool to investigate issues in astrophysics, high-energy physics, and fundamental problems in gravity. It is often complementary to fully-fledged nonlinear evolutions and instrumental to interpret some results of numerical simulations. Several modern applications require advanced tools to investigate the linear dynamics of generic small perturbations around stationary black holes. Here, we present an overview of these applications and introduce extensions of the standard semianalytical methods to construct and solve the linearized field equations in curved spacetime. Current state-of-the-art techniques are pedagogically explained and exciting open problems are presented.
\keywords{Black holes; Perturbation theory; Gravitational waves; Numerical methods.}
\end{abstract}

\ccode{PACS numbers:04.25.Nx;04.30.-w;04.50.Kd;04.70.-s}

\section{A perturbative approach to BH dynamics}\label{sec:intro}	
The scope of these notes is to introduce some state-of-the-art tools to investigate the dynamics of small perturbations around stationary and axisymmetric black holes (BHs) at linear level. A perturbative analysis of BH dynamics is crucial in several contexts, ranging from astrophysics to high-energy physics. There exists a number of excellent reviews on the subject~\cite{Nollert:1999ji,Kokkotas:1999bd,Ferrari:2007dd,Berti:2009kk,Konoplya:2011qq,Cardoso:2012qm} to which we refer for details and for an exhaustive account of the literature. The stability analysis of BH spacetimes, BH ringdown after binary mergers, gravitational-wave emission in astrophysical processes and even the gravity/gauge correspondence are just the most noteworthy contexts in which BH perturbation theory is relevant. The problem is --~directly or indirectly~-- reduced to solving the linearized dynamics of some fields on a curved background. Hence, we wish to address the following question:
\emph{``How do small perturbations propagate on the background of a stationary BH?''}

The dawn of BH linear perturbation theory dates back to 1957 due to the pioneering work by Regge and Wheeler. During the BH Golden Age (1963--1973) the field experienced a tremendous boost thanks to the fundamental contributions by Zerilli, Vishveshwara, Teukolsky and Press (original references can be found in the reviews mentioned above). Already in 1973, the master equations governing the massless perturbations of the Kerr metric were known for scalar, electromagnetic and  gravitational perturbations and they have been more recently extended to include fields of spin $1/2$ and $5/2$. 

The great advantage of Teukolsky's equations is that the angular dependence
has been completely separated by a suitable choice of the angular basis, written in terms of spheroidal harmonics. Therefore, the remaining equations only contain a radial and a time dependence: the problem is reduced to a $1+1$ evolution in the time domain or, due to the stationarity of the background, to a simple one-dimensional problem in the frequency domain.

Over the years several numerical methods have been implemented in order to solve the master equations subjected to some initial and boundary conditions in the time domain and to physically-motivated boundary conditions in the frequency domain. In the latter case, the equations reduce to a one-dimensional eigenvalue problem. Imposing boundary conditions at the BH horizon and at infinity singles out an infinite~\cite{Kokkotas:1999bd} number of complex frequencies, $\omega=\omega_R+i\omega_I$. The nonzero imaginary part of the modes is due to dissipation, both at infinity (because of the emission of gravitational waves) and at the horizon (which behave as a one-way, viscous membrane). An important class of eigenfrequencies are the so-called BH quasi-normal modes (QNMs) which are thoroughly discussed in the reviews above and that we shall also discuss in some detail.

In the frequency domain, the most popular techniques to compute the BH eigenfrequencies include: WKB approximations,  highly-efficient continued-fraction techniques, series solutions for asymptotically Anti de Sitter (AdS) BHs, Breit-Wigner resonance method for long-lived modes and monodromy techniques [see also~\cite{Fiziev:2012mw} for other approaches]. The spectrum of spinning BHs is extremely rich and each of these methods is best-suited to explore some specific region~\cite{Berti:2009kk}.

Nevertheless Teukolsky's approach --~based on a Newman-Penrose tetrad decomposition in terms of the components of the Weyl tensor~\cite{Chandra}~-- is limited to cases in which the angular dependence is separable. This is a fairly restrictive requirement, because the Kerr metric in four dimensions represents an exception in this regard. Indeed, separability usually requires that the background spacetime enjoys special symmetries. The underlying property that, at least in some cases~\footnote{The Myers-Perry metric is the generalization of the Kerr solution to higher dimensions. Even if this spacetime is Type D in the Petrov classification, separability in the Teukolsky formalism is still an open problem in the general case, see the main text. Another notable example of nonseparability in Type-D background is the case of massive spin-1 perturbations on a Kerr metric. This is discussed later on.} allows for the exceptional separability of the perturbation equations on a Kerr spacetime is the fact that the latter is of Petrov type-D. 

In recent times, it has become clear that the standard Teukolsky's approach is inadequate to deal with more generic classes of background metrics, which naturally emerge in a variety of applications. We list here the most noteworthy ones:
\begin{itemize}
 \item \textbf{Tests of the no-hair theorem.} Uniqueness theorems in general relativity (GR)  guarantee that a stationary BH is necessarily axisymmetric and described by the Kerr-Newman metric. This is not generically the case in modified theories of gravity, whose spinning BH solutions deviate parametrically from their GR counterparts. Furthermore, even in those theories which share the same BH solutions as in GR~\cite{Sotiriou:2011dz}, the dynamics of linear perturbations is different~\cite{Barausse:2008xv} and encodes information of the underlying theory. Near-future gravitational-wave observations will probe regions of strong gravitational field (e.g. by detecting the signal from a BH-binary merger or from the inspiral of small compact objects around supermassive BHs~\cite{AmaroSeoane:2012km}) and are in principle able to detect deviations from the Kerr solution. This, however, requires (at the very least) to understand the gravitational-wave emission from nonKerr BHs in alternative theories and Teukolsky's approach does not seem adequate in this case.
 \item \textbf{Spinning BHs in higher-dimensions.} Uniqueness theorems do not extend to dimensions $D$ higher than four and multiple spinning black objects with the same asymptotic charges are known when $D>4$. In addition, several axes of rotation exist and, correspondingly, these objects are characterized by multiple angular momenta. The phase diagram depends on the number of dimensions and on the angular momenta and it generically shows bifurcation points and phase transitions. Correspondingly, several spinning objects in higher dimensions are unstable~\cite{Shibata:2010wz,Dias:2009iu} and their linear stability analysis is still an open question in a generic setup. Furthermore, in a semiclassical treatment of BH evaporation, the calculation of greybody factors (which may be of direct interest for ongoing experiments~\cite{Cardoso:2012qm}) relies on our ability to understand wave scattering in rotating BH spacetimes. Extensions of Teukolsky's approach to dimensions higher than four are both challenging and of great relevance.
 \item \textbf{Kerr-Newman BHs in GR.} Despite the 40-year-long effort, gravito-electromagnetic perturbations on a Kerr-Newman metric do not appear to be separable in the standard Teukolsky's formalism~\cite{Chandra} [see Ref.~\cite{Pani:2013ija,Pani:2013wsa} for a recent attempt in the slowly-rotation approximation discussed below]. This is highly disappointing because, as mentioned above, the Kerr-Newman metric describes the most general stationary BH solution in GR and it is remarkably simple, being defined by three parameters only: the mass, the spin and the electric charge. 
 \item \textbf{Massive bosonic perturbations of a Kerr BH.} Interestingly, not all probe-field perturbations are separable even in a Kerr background. Massive perturbations of spin equal or greater than one do not appear to be separable in the standard approach. Besides their theoretical interest \emph{per s\'e}, light massive bosonic fields around spinning BHs give rise to interesting effects~\cite{Arvanitaki:2009fg,Cardoso:2011xi} which can be revealed by a linearized analysis. These fields are ubiquitous in extensions of the standard model, for instance in the so-called axiverse scenario~\cite{Arvanitaki:2009fg} or in models describing light vector fields and massive gravitons~\cite{Babichev:2013una,Brito:2013wya}. Astrophysical signatures of the dynamics of these fields around BHs may open new windows to test particle physics beyond the standard model.
 \item \textbf{Astrophysical BHs.} Realistic BHs that are formed as end-states of sufficiently massive stars are surrounded by matter. The prototypical example are accretion disks, but spinning BHs are also believed to host magnetic fields and give rise to jet emissions. These configurations are typically dynamical and not particularly symmetric but, even when they can be approximately treated as stationary and axisymmetric, their gravitational perturbations are coupled to those of the surrounding matter, requiring some extension of the standard approach.	
 \item \textbf{BHs in the gravity/gauge duality.} Last but not least, all previous considerations about the challenge of studying linear perturbations in nonKerr spacetimes apply to the case of asymptotically AdS BHs. These solutions are of great relevance in the so-called gravity/gauge duality and in its phenomenological applications to strongly-coupled condensed-matter systems. In the correspondence, some correlation functions and transport coefficients of the dual holographic theory are related to the lowest order BH QNMs and to the BH linear response in general~\cite{Hartnoll:2009sz,Berti:2009kk,Pani:2012zz}. In this context, BHs endowed with nontrivial (scalar, electromagnetic, nonAbelian and fermionic) fields are usually considered, and understanding the thermalization processes in the dual theory relies on the ability of solving the linear dynamics on the hairy background. 
\end{itemize}
Of course, in most cases listed above a linear analysis cannot be conclusive and must be complemented with exact solutions and extended by full-fledged numerical evolutions. The latter however, would greatly benefit by a detailed linearized analysis. The two approaches are often complementary to each other and have their own disjoint domain of validity.

The number of interesting applications that require advanced tools in BH perturbation theory is a good predictor of the relevance of the topics we are going to discuss and of the exciting time lying ahead those who will embrace this field. 

\subsubsection*{Notation}
Hereafter Greek indices stand for spacetime coordinates. Capital Latin indices are used to denote nonangular coordinates, whereas lower-case Latin indices at the beginning of the alphabet (e.g. $a,b,...$) denote angular coordinates.  Latin indices in the middle of the alphabet (e.g. $i,j,k,n,I,L...$) denote unspecified indices (e.g. matrix indices). According to this notation, the four-dimensional coordinate vector reads $x^\mu=(y^A,z^a)$ with $y^A=(t,r)$ and $z^a=(\vartheta,\varphi)$. Unless otherwise stated, we adopt natural units $\hbar=G=c=1$.

\section{Perturbations of nonspinning BHs}	\label{sec:nonspinning}
We start by discussing the simpler case of nonrotating BHs. We consider static and spherically symmetric spacetimes in four dimensions, although in the nonspinning case most of the discussion can be easily extended to higher dimensions and to other topologies (e.g. to higher-dimensional black branes). The line element reads
\begin{equation}
 ds^2\equiv g_{\mu\nu}dx^\mu dx^\nu=-F(r)dt^2+B(r)^{-1}dr^2+r^2 d\Omega^2+\delta g_{\mu\nu}dx^\mu dx^\nu\,,\label{metric}
\end{equation}
where $F(r)$ and $B(r)$ are background quantities that depend on the specific solution and $\delta g_{\mu\nu}$ are first order terms. Our goal in this section is to derive the perturbation equations for $\delta g_{\mu\nu}$ and solve them numerically in a quite generic class of problems. In doing so, we shall present several methods that can be directly adapted to study perturbations of rotating metrics.

\subsection{Harmonic decomposition}
In order to derive the perturbation equations, we follow a standard decomposition of the metric and possible other fields in tensor spherical harmonics~\cite{Chandra}.
The decomposition of the metric is based on the transformation properties of the ten components of the perturbation tensor $\delta g_{\mu\nu}$ under a rotation of the 
frame around the origin. When considered as covariant 
quantities on the sphere, they transform as three $SO(2)$ scalars $\delta g_{AB}$, two $SO(2)$ vectors $\delta g_{Aa}$ and 
one $SO(2)$ second--order tensor $\delta g_{ab}$ and they can be expanded in the complete basis 
constituted by the spherical harmonics of different rank. 

Furthermore, perturbations naturally divide into two classes, accordingly to their transformation properties under parity, namely
\begin{equation}
 \delta g_{\mu\nu}(t,r,\vartheta,\varphi)=\delta g_{\mu\nu}^{\rm odd}(t,r,\vartheta,\varphi)+\delta g_{\mu\nu}^{\rm even}(t	,r,\vartheta,\varphi)
\end{equation}
with
\begin{equation}\label{oddpart}
\delta g_{\mu\nu}^{\rm odd} =
 \begin{pmatrix}
  0 & 0 & h_0^\ell S_\vartheta^{\ell} & h_0^\ell S_\vartheta^{\ell} \\
  * & 0 & h_1^\ell S_\vartheta^{\ell} & h_1^\ell S_\vartheta^{\ell} \\
  *  & *  & -h_2^\ell\frac{X^\ell}{\sin\vartheta} & h_2^\ell\sin\vartheta W^\ell  \\
  * & * & * & h_2^\ell\sin\vartheta X^\ell
 \end{pmatrix}\,,
\end{equation}
\begin{equation}\label{evenpart}
\delta g_{\mu\nu}^{\rm even}=
\begin{pmatrix}
g_{tt}^{(0)} H_0^\ell Y^\ell & H_1^\ell Y^\ell & \eta_0^\ell Y_{,\vartheta}^\ell& \eta_0^\ell Y_{,\varphi}^\ell\\
  * & g_{rr}^{(0)} H_2^\ell Y^\ell & \eta_1^\ell Y_{,\vartheta}^\ell & \eta_1^\ell Y_{,\varphi}^\ell\\
  *  & *  & r^2\left[K^\ell Y^\ell+G^\ell W^\ell\right] & r^2  G^\ell X^\ell  \\
  * & * & * & r^2\sin^2\vartheta\left[K^\ell Y^\ell-G^\ell W^\ell\right]
 \end{pmatrix}\,.
\end{equation}
%
where asterisks represent symmetric components, $Y^{\ell}=Y^{\ell}(\vartheta,\varphi)$ are the scalar spherical harmonics and we have defined
\begin{eqnarray}
(S_\vartheta^{\ell},S_\varphi^{\ell})&\equiv&\left(-\frac{Y^{\ell}_{,\varphi}}{\sin\vartheta}
,\sin\vartheta Y^{\ell}_{,\vartheta}\right)\,.\\
(X^{\ell},W^{\ell})&\equiv&\left(2(Y^{\ell}_{,\vartheta\varphi}-\cot\vartheta Y^{\ell}_{,\varphi}),Y^{\ell}_{,\vartheta\vartheta}-\cot\vartheta Y^{\ell}_{,\vartheta}-\frac{Y^{\ell}_{,\varphi\varphi}}{\sin^2\vartheta}\right)\,. \label{XW}
\end{eqnarray}
Here and in the following, a sum over the harmonic indices $\ell$ and $m$ (such that $|m|\leq\ell$) is implicit\footnote{Furthermore, from now on we will append the relevant   multipolar index $\ell$ to any perturbation variable but we will omit the index $m$, because in an axisymmetric background it is possible to decouple the perturbation equations so that all quantities have the same value of $m$.}.
Under parity transformations ($\vartheta\rightarrow\pi-\vartheta$,
$\varphi\rightarrow\varphi+\pi$): polar and axial perturbations are
multiplied by $(-1)^\ell$ and $(-1)^{\ell+1}$, respectively.
The odd and even sectors are also referred to as ``axial'' and ``polar'' and we shall use the two notations indistinctly.
The functions $(H_0,H_1,H_2,K,G,\eta_0,\eta_1)^{\ell}$ and $(h_0,h_1,h_2)^{\ell}$ only depend on $t$ and $r$ and describe the polar parity metric
perturbations and the axial parity metric perturbations, respectively. 

Depending on the number of polarizations of the graviton, there can be a residual gauge freedom in the metric perturbations that can be used to simplify the equations. For a massless graviton it is convenient to adopt the so-called Regge-Wheeler gauge, in which $\eta_i^\ell\equiv G^\ell\equiv h_2^\ell\equiv0$. In this gauge, we are then left with four polar functions and two axial functions. However, in modified theories of gravity the graviton can propagate more than two polarizations. For example a massive graviton propagates five degrees of freedom and there is no residual gauge freedom in the expansion above.

Finally, in presence of other fundamental fields, we decompose them in spherical harmonics of the corresponding type. Vector fields are decomposed in a basis of vector spherical harmonics, whereas scalar fields are decomposed in scalar spherical harmonics. This procedure is very general and can be performed in any spherically symmetric spacetime. Noteworthy, in this decomposition axial and polar perturbations belong to two separate sets of equations and also perturbations with different harmonic index $\ell$ are separated. For a given $\ell$ we are then left with two systems of equations, one for the axial sector and one for the polar sector, which completely characterize the linear response of the system.

\subsection{Computing the eigenfrequencies}\label{sec:computingQNMs}
Typically, for a given $\ell$, the axial and polar sectors can be separately written as a coupled system of the form\footnote{Equation~\eqref{systemt} can be replaced by a generic system of ordinary differential equations which is of second-order in time and of second-order in the radial coordinate. The rest of the discussion would be very similar to that given in the text. We find it convenient to use Eq.~\eqref{systemt}, though, mostly to simplify the notation.}:
\begin{equation}
 \left[-\frac{d^2}{dt}+\frac{d^2}{dr_*^2}\right]\mathbf{Y}-\mathbf{V}(r)\mathbf{Y}=0\,. \label{systemt}
\end{equation}
where $r_*$ are some suitable coordinate (we assume $r_*\to-\infty$ as $r\to r_+$ and $r_*\to\infty$ as $r\to \infty$), $\mathbf{Y}$ is a $N$-vector and $\mathbf{V}$ is a $N\times N$ matrix, which depends only on $r$ and $\ell$ and not on $t$ and $m$ if the background is stationary and spherically symmetric. It is often convenient to Fourier-transform to the frequency domain. By defining $\mathbf{Y}=\int dt e^{-i\omega t} \mathbf{\tilde{Y}}$, we get
\begin{equation}
 \left[\frac{d^2}{dr_*^2}+\omega^2-\mathbf{V}(r)\right]\mathbf{\tilde{Y}}=0\,. \label{system}
\end{equation}
For brevity, in the rest of this section we omit the tilde, but all quantities have to be understood as Fourier transforms. 
Furthermore, we assume $V_{ij}\to0$ at the BH outer horizon, $r\to r_+$, and $V_{ij}\to \mu^2 \delta_{ij}$ at infinity, $r\to \infty$. The latter is the typical behavior of the potential for massive fields and it reduces to the more common massless case when $\mu=0$. Possible generalizations to different classes of potentials are straightforward and left for exercise. The case of asymptotically (A)dS spacetime, in which $V_{ij}\to\infty$ at infinity is discussed in the next sections.

When physically motivated boundary conditions at the horizon and at infinity are imposed, the system~\eqref{system} forms an eigenvalue problem for the frequency $\omega$. Our goal in this section is to compute the eigenfrequency spectrum.

Close to the horizon, the solution behaves as a superposition of ingoing and outgoing waves and physical boundary conditions require a purely ingoing-wave condition~\cite{Berti:2009kk}. Therefore, the desired behavior of the solution close to the horizon reads:
\begin{equation}
  Y_i\sim e^{-i \omega r_*}\sum_n b_n^{(i)}(r-r_+)^n \qquad r\to r_+\,,\label{series_hor}
\end{equation}
where $n>0$, $Y_i$ is the $i$th-component of $\mathbf{Y}$ and the coefficients $b_n^{(i)}$
can be computed in terms of $b_0^{(i)}$ by solving the near-horizon equations
order by order. 
The general asymptotic behavior at infinity reads:
\begin{equation}
 Y_i\sim B_{(i)} e^{-k_\infty r_*}
+C_{(i)} e^{k_\infty r_*} \qquad r\to \infty\,,\label{BCinf}
\end{equation}
where $k_\infty=\sqrt{\mu^2-\omega^2}$ and, without loss of generality, we choose the root such that $\rm{Re}[k_\infty]>0$.
The boundary conditions $B_{(i)}=0$ define purely outgoing waves at
infinity, i.e. QNMs~\cite{Berti:2009kk}. In the case of massive perturbations the condition
$C_{(i)}=0$ is also allowed and physically motivated. The latter defines states which are spatially localized within the
vicinity of the BH and decay exponentially at spatial infinity,
i.e. bound states~\cite{Dolan:2007mj,Rosa:2011my}. In fact, if such modes exist in a BH spacetime they are ``quasi'' bound because, even if they do not propagate energy to infinity, they dissipate energy at the BH event horizon. Dissipation at the horizon allows for interesting effects related to the superradiance of spinning BH spacetimes~\cite{Teukolsky:1974yv,Cardoso:2012zn,Cardoso:2011xi} and may also produce instabilities~\cite{Detweiler:1980uk,Pani:2012vp}, whose timescale $1/\omega_I$ can be computed within the linearized approximation.

\subsubsection{Matrix-valued continued-fraction method}
Since the seminal work by Leaver~\cite{Leaver:1985ax}, it is well-known that many classes of eigenvalue problems in GR can be solved through continued-fraction techniques. This is a highly-efficient method which is well-suited to Schroedinger-like potentials that contain only (fractions of) powers of $1/r$. In this case, the eigenfunction can be written as a series whose coefficients satisfy a finite-term recurrence relation. A robust method to solve three-term recurrence relations is available and any higher-order recurrence relation can be reduced to a three-term one via Gaussian elimination~\cite{Berti:2009kk}. The efficiency of this method makes it one of the optimal tools to solve linear eigenvalues problems.
Here, we discuss a generalization of the method, to solve coupled systems of equations in the form~\eqref{system}~\cite{Rosa:2011my,PhysRevE.59.5344}.

In order to optimize the recurrence relation, it is important to choose a suitable ansatz for the eigenfunctions. Let us consider the case in which the background solution has a single horizon $r_+$, such that $F(r_+)=0$\footnote{In case of multiple horizons a slightly different ansatz is more convenient. See e.g.~\cite{Leaver:1990zz,Berti:2005eb}.}. Then a convenient ansatz for the solution of the system~\eqref{system} reads:
\begin{equation}
 Y_i=e^{-i\omega r_*}r^{-\nu} e^{qr} \sum_n a_n^{(i)}F(r)^n
\end{equation}
where $\nu$ is a constant that depends on the specific problem and $q=\pm\sqrt{\mu^2-\omega^2}$. In the case of massive fields, $\mu\neq0$, the sign of the real part of $q$ selects the correct boundary condition at infinity: the plus sign refers to QN frequencies, whereas the minus sign refers to quasi-bound states.
Inserting the equation above into Eq.~\eqref{system}, it is possible to obtain a recurrence relation for the vectors $\mathbf{a}_n$. Let us start with a simple case and assume that the system reduces to a three-term matrix-valued recurrence relation:
\begin{eqnarray}
&&\boldsymbol{\alpha}_0 \mathbf{a}_{1} + \boldsymbol{\beta}_0 \mathbf{a}_{0} = 0\qquad n=0\,, \label{recurrence0}\\
&&\boldsymbol{\alpha}_n \mathbf{a}_{n+1} + \boldsymbol{\beta}_n \mathbf{a}_{n} + \boldsymbol{\gamma}_n \mathbf{a}_{n-1} = 0\,, \qquad  n > 0\,,\label{recurrencen}
\end{eqnarray}
The matrices $\boldsymbol{\alpha}_n$, $\boldsymbol{\beta}_n$ and $\boldsymbol{\gamma}_n$ are generically nondiagonal for coupled systems. 
In this case, a three-term recurrence relation as the one above can be solved in the following way. First, we define the ladder matrix $\mathbf{R}_n^+$ such that
\begin{equation}
 \mathbf{a}_{n+1}=\mathbf{R}_n^+ \mathbf{a}_n\,.
\end{equation}
Taking Eq.~\eqref{recurrencen} with $n\to n+1$, solving for $\mathbf{a}_{n+1}$ and using the equation above, we obtain:
\begin{equation}
   \mathbf{R}_n^+=-\left[\boldsymbol{\beta}_{n+1}+\boldsymbol{\alpha}_{n+1} \mathbf{R}_{n+1}^+\right]^{-1}\boldsymbol{\gamma}_{n+1}\,. \label{Rn}
\end{equation}
Finally, the recurrence relation is solved by imposing Eq.~\eqref{recurrence0}, i.e.
by searching for roots of the equation $\mathbf{M}\mathbf{a}_0=0$, with
\begin{equation}
 \mathbf{M}\equiv \boldsymbol{\beta_0}-\boldsymbol{\alpha}_0\left[\boldsymbol{\beta}_{1}-\boldsymbol{\alpha}_{1}(\boldsymbol{\beta}_2+\boldsymbol{\alpha}_2\mathbf{R}_{2}^+) \boldsymbol{\gamma}_2\right]^{-1}\boldsymbol{\gamma}_1\,,
\end{equation}
where $\mathbf{R}_{2}^+$ is obtained recursively from Eq.~\eqref{Rn}. Therefore, for nontrivial solutions the eigenfrequencies are the roots of the determinant:
\begin{equation}
 {\rm det}\mathbf{M}=0\,.
\end{equation}
In practice, one usually fixes a large truncation order $N$ and initializes $\mathbf{R}_N^+$ arbitrarily. Then, Eq.~\eqref{Rn} is used to obtain $\mathbf{R}_{N-1}^+$, $\mathbf{R}_{N-2}^+$ and so on, down to $\mathbf{R}_2^+$. After this cascade of matrix-inversions, the matrix $\mathbf{M}$ can be constructed. Clearly, convergence of the results for different values of $N$ must be verified a posteriori.

This procedure might appear a bit abstract at first sight, but it is indeed straightforward to implement. In the notebook \url{CF_matrix_3terms.nb}~\cite{webpage}, we present a short implementation of the matrix-valued continued-fraction method to compute scalar, electromagnetic and gravitational modes of a Schwarzschild BH in GR. Since in this case the equations are decoupled, the matrices defining the recurrence relation are diagonal. Note that, with a few lines of code, it is possible to compute the modes of perturbations of different spin in a single step.

Let us consider the case in which the recurrence relation involves more than three terms. As an example, we consider a four-term recurrence relation:
\begin{eqnarray}
&&\boldsymbol{\alpha}_0 \mathbf{a}_{1} + \boldsymbol{\beta}_0 \mathbf{a}_{0} = 0\,, \quad \hspace{5cm} n=0\,,\nn\\
&&\boldsymbol{\alpha}_1 \mathbf{a}_{2} + \boldsymbol{\beta}_1 \mathbf{a}_{1} + \boldsymbol{\gamma}_1 \mathbf{a}_{0} = 0\,, \quad  \hspace{3.85cm} n =1\,,\nn \\
&&\boldsymbol{\alpha}_n \mathbf{a}_{n+1} + \boldsymbol{\beta}_n \mathbf{a}_{n} + \boldsymbol{\gamma}_n \mathbf{a}_{n-1} + \boldsymbol{\delta}_n \mathbf{a}_{n-2} = 0\,, \quad \hspace{1.375cm} n > 1\,,\nn 
\end{eqnarray}
where $\mathbf{a}_n$ is a $N$-dimensional
vector and $\boldsymbol{\alpha}_n$,
$\boldsymbol{\beta}_n$, $\boldsymbol{\gamma}_n$,
$\boldsymbol{\delta}_n$, $\boldsymbol{\rho}_n$ and
$\boldsymbol{\sigma}_n$ are $N\times N$ \emph{invertible} matrices.
The order of the recurrence relation can be reduced by using a matrix-valued version of the Gaussian
elimination~\cite{Leaver:1990zz,Berti:2009kk}. By defining 
\begin{eqnarray}
 \boldsymbol{\tilde\alpha}_n &=& \boldsymbol{\alpha}_n \,,\\
  \boldsymbol{\tilde\beta}_0 &=& \boldsymbol{\beta}_0 \,,\\
    \boldsymbol{\tilde\gamma}_0 &=& \boldsymbol{\gamma}_0 \,,\\
   \boldsymbol{\tilde\beta}_n &=& \boldsymbol{\beta}_n-\boldsymbol{\delta}_n\left[\boldsymbol{\tilde\gamma}_{n-1} \boldsymbol{\tilde\alpha}_{n-1}\right]^{-1} \quad n>0\,, \\
      \boldsymbol{\tilde\gamma}_n &=& \boldsymbol{\gamma}_n-\boldsymbol{\delta}_n\left[\boldsymbol{\tilde\gamma}_{n-1} \boldsymbol{\tilde\beta}_{n-1}\right]^{-1} \quad n>0\,,
\end{eqnarray}
the tilded matrices satisfy the same three-term recurrence relation as~\eqref{recurrence0}--\eqref{recurrencen}. This procedure can be extended to reduce any matrix-value recurrence relation (provided some matrices are invertible) to a thee-term one\footnote{\textbf{Exercise:} Extend the code in \url{CF_matrix_3terms.nb} in order to reduce a generic $N$-term recurrence relation into a three-term one.}, which can be solved as explained above.

\subsubsection{Matrix-valued direct integration}
It is possible to compute the characteristic frequencies of the system~\eqref{system} also using a direct
integration shooting method~\cite{Ferrari:2007rc,Rosa:2011my,Pani:2012bp}. The idea is to integrate the system from the horizon with boundary conditions~\eqref{series_hor} outwards to infinity, where we impose either $B_{(i)}=0$ or $C_{(i)}=0$, depending on the physical problem at hand. The procedure is explained here in general and an example is given in the notebook \url{DCS_DI.nb}~\cite{webpage}, where we compute the QNMs of a Schwarzschild BH in Dynamical Chern-Simons (DCS) gravity [see also Section~\ref{sec:example} below] with this method.

Let us start with a system of $N$ second-order ordinary differential equations (ODEs) for $N$ perturbation functions as in Eq.~\eqref{system}. Starting with a near-horizon solution as $\eqref{series_hor}$, a family of solutions at infinity is then characterized by $N$
parameters, corresponding to the $N$-dimensional vector of the near-horizon
coefficients, $\mathbf{b_0}=\{b_0^{(i)}\}$ ($i=1,...,N$).
At infinity we look either for
exponentially decaying solutions, $C_{(i)}=0$, or for QNMs, $B_{(i)}=0$.
In both cases, the spectrum can be obtained as follows. We first choose a suitable orthogonal basis
for the $N$-dimensional space of the initial coefficients $b_0^{(i)}$. We perform $N$ integrations from the horizon to
infinity and construct the $N\times N$ matrix:
\begin{eqnarray}
 \label{matrix_coupled}
\mathbf{S}(\omega)=
\begin{pmatrix}
A_1^{(1)} & A_1^{(2)}   & ... & A_1^{(N)}\\ 
A_2^{(1)} & A_2^{(2)}   & ... & ... \\
... & ...   & ... & ... \\
... & ...   & ... & ... \\
A_N^{(1)} & ...   & ... & A_N^{(N)} \\
\end{pmatrix}\,,
\end{eqnarray}
where $A\equiv B$ if we want to compute QNMs, whereas $A\equiv C$ if we want to compute quasi-bound states, respectively (of course, mixed boundary conditions are possible). The superscripts denote a particular vector of the basis,
i.e. $A_{i}^{(1)}$ corresponds to $\mathbf{b_0}=\{1,0,0,...,0\}$,
$A_i^{(2)}$ corresponds to $\mathbf{b_0}=\{0,1,0,...,0\}$ and
$A_i^{(N)}$ corresponds to $\mathbf{b_0}=\{0,0,0,...,1\}$.
Finally, the characteristic frequencies
$\omega_0=\omega_R+i\omega_I$ are obtained by imposing
\begin{equation}
 \det{\mathbf{S}(\omega_0)}=0\,.\label{det_modes}
\end{equation}
To summarize, by performing $N$ integrations from the horizon to
infinity we can construct the single-valued complex function $\det{\mathbf{S}(\omega_0)}$ and the problem of finding the eigenfrequencies is reduced to finding the complex roots of this function. This can be implemented, for instance, by a simple one-parameter shooting method [cf. notebook \url{DCS_DI.nb}~\cite{webpage} for an example].

Note that a direct integration performs extremely well to compute quasi-bound state modes, because in this case the condition $C_{i}=0$ can be imposed from the leading behavior of the fields at infinity. In many cases, the accuracy of the results may exceed that achievable by continued-fraction techniques, whose convergence properties deteriorate in some case (e.g. for ultra-slowly-damped modes).  On the other hand, the condition $B_{(i)}=0$ requires to extract the subdominant, exponentially suppressed behavior at large distance and this can be contaminated by numerical errors. 
This makes the direct-integration approach nonoptimally suited to compute QNMs. Nonetheless, if the imaginary part of the mode is sufficiently small compared to the real part, precise results can be obtained by integrating up to moderately large values of $r$ and including higher-order terms in the series expansion~\eqref{BCinf} at infinity to reduce truncation errors. Typically this allows to compute the fundamental mode and possibly the first few overtones. In spite of this limitation, the direct integration technique is extremely flexible, because it does not rely on any particular property of the matrix-valued potential $\mathbf{V}(r)$ and can be applied to essentially any class of boundary value problems.

\subsubsection{Breit-Wigner resonance method}
When dealing with complicated systems of coupled equations, the direct integration discussed above can be time demanding.
In cases in which the eigenfrequency spectrum supports slowly
damped modes, i.e. those with $\omega_I\ll\omega_R$, we can adopt an approximate procedure known as Breit-Wigner resonance method, or the standing-wave
approach~\cite{1969ApJ...158..997T,Chandrasekhar:1992ey,Ferrari:2007rc,Berti:2009wx}. 
As we now discuss, the great advantage of this method is that the eigenvalue problem can be solved by looking for minima of a
real-valued function of a real variable~\cite{Chandrasekhar:1992ey}.

By expanding Eq.~\eqref{det_modes} about $\omega_R$ and assuming $\omega_I\ll
\omega_R$ we get~\cite{Ferrari:2007rc}
\begin{equation}
\det{\mathbf{S}(\omega_0)}\simeq\det{\mathbf{S}(\omega_R)}
+i\omega_I\left.\frac{d\left[\det{\mathbf{S}(\omega)}\right]}{d\omega}\right|_{\omega_R}=0\,.
\end{equation}
We consider the function $\det{\mathbf{S}}$
restricted to real values of $\omega$.
Using the relation above, a Taylor expansion for real $\omega$ close to $\omega_R$ yields:
\begin{equation}
\det{\mathbf{S}(\omega)}\simeq\det{\mathbf{S}
(\omega_R)}\left[1-\frac{\omega-\omega_R}{i\omega_I}
\right]\propto\omega-\omega_R-i\omega_I\,.
\end{equation}
Therefore, in the region of the real--$\omega$ axis close to the real part of the mode, we have
\begin{equation}
 |\det{\mathbf{S}\left(\omega\right)}|^2\propto
\left(\omega-\omega_R\right)^2+\omega_I^2\,,\label{parabola}
\end{equation}
that is, the function $|\det{\mathbf{S}}|^2$ is simply a parabola when $\omega\approx\omega_R$.
To summarize, to find the slowly-damped modes it is sufficient to integrate the
system~\eqref{system} $N$ times for real values of the frequency $\omega$, construct
the matrix ${\mathbf{S}\left(\omega\right)}$ and find the minima of the
function $|\det{\mathbf{S}}|^2$, which represent the real part of the
modes. Then the imaginary part (in modulus) of the mode can be extracted through
a quadratic fit, as in Eq.~\eqref{parabola}.
We postpone an application of the Breit-Wigner method to the case of slowly-rotating BHs discussed below, in which the great efficiency of this method becomes evident.
%

\subsubsection{Matrix-valued series method}
The methods discussed so far can be implemented in asymptotically flat spacetimes (and with some minor modification in asymptotically de Sitter spacetimes). However, computing the eigenfrequencies of BHs in asymptotically AdS spacetime usually requires a separate treatment, due to the different behavior of the fields at the AdS boundary. When the problem is described by a single second-order ODE, a series method~\cite{Horowitz:1999jd} proves to be very efficient.
In this method, local solutions near the regular singular points (at the horizon and at spatial infinity) are represented in terms of convergent Frobenius series. In various cases of interest, the radius of convergence of the series is equal to or larger than the
interval of interest. This is the case of large spherically symmetric BHs (i.e. $r_+\gg L$, $L$ being the AdS radius) or of black branes. 
On the other hand, for small BHs ($r_+\ll L$) the convergence properties of the series are very poor. In such cases, if the spectrum supports slowly-damped modes, these can be computed by the Breit-Wigner method discussed above~\cite{Berti:2009wx}.
Since large AdS BHs and black branes are relevant in the context of the gauge/gravity duality~\cite{Hartnoll:2009sz}, here we discuss an extension of the series method to deal with coupled systems which arise quite naturally in the case of hairy AdS BHs~\cite{Delsate:2011qp}.

We consider a coupled system of $N$ equations as in Eq.~\eqref{system}. The near-horizon behavior of the solution is given by Eq.~\eqref{series_hor} but, at variance with the asymptotically flat case, the equations present a regular singularity at spatial infinity. Generically, near spatial infinity the solution behaves as
\begin{equation}
 Y_i\to A_i r^{\alpha_i}+B_i r^{\beta_i} \,, \label{BCinfAdS}
\end{equation}
where $\alpha_i$ and $\beta_i$ depend on the specific problem at hand. Typically, $\alpha_i\beta_i<0$, so that only one of the two terms above is regular. By imposing that the coefficient of the irregular term vanishes (Dirichlet boundary conditions) the eigenvalue problem is specified. In some cases, for example for gravitational perturbations of a Schwarzschild-AdS BH, both terms in Eq.~\eqref{BCinfAdS} are regular and several inequivalent choices of the boundary conditions are possible. In the context of the AdS/CFT correspondence, Robin boundary conditions have been also considered~\cite{Moss:2001ga,Dias:2013sdc}. Here for simplicity we focus on Dirichlet boundary conditions, i.e. $Y_i\to0$ at spatial infinity.

By defining a new variable $x=1/r$, and factorizing the near-horizon behavior $Y_i(x)=e^{-i\omega r_*}Z_i(x)$, the system of equations can be written in the form:
\begin{equation}
 (x-x_+)s(x)\frac{d^2\mathbf{Z}}{dx^2}+ t(x)\frac{d\mathbf{Z}}{dx}+\frac{\mathbf{u(x)}}{x-x_+}\mathbf{Z}=0\,,
\end{equation}
where $x_+=1/r_+$ and $\mathbf{u}(x)$ is a matrix related to $\mathbf{V}(r)$. The method consists in finding a local Frobenius solution near the singular point $x=x_+$,
\begin{equation}
 Z_i=(x-x_+)^{\gamma_{i}}\sum_{n=0}^\infty a_n^{(i)}(\omega) (x-x_+)^n\,, \label{Frobenius}
\end{equation}
where $\gamma_i$ depend on the specific problem. The series coefficients $a_n^{(i)}$ can be computed iteratively and they only depend on the $N$-dimensional vector $\mathbf{a_0}\equiv\left\{a_0^{(i)}\right\}$. 

As discussed in the case of direct integration, we can choose a suitable orthogonal basis
for the $N$-dimensional space of the initial coefficients $a_0^{(i)}$. For each element of the basis, we construct the $N\times N$ matrix:
\begin{eqnarray}
 \label{matrix_coupledZ}
\mathbf{S}(\omega)=\lim_{r\to\infty}
\begin{pmatrix}
Z_1^{(1)} & Z_1^{(2)}   & ... & Z_1^{(N)}\\ 
Z_2^{(1)} & Z_2^{(2)}   & ... & ... \\
... & ...   & ... & ... \\
... & ...   & ... & ... \\
Z_N^{(1)} & ...   & ... & Z_N^{(N)} \\
\end{pmatrix}\,,
\end{eqnarray}
where again the superscripts denote a particular vector of the basis,
i.e. $Z_{i}^{(1)}$ corresponds to $\mathbf{a_0}=\{1,0,0,...,0\}$,
$Z_1^{(2)}$ corresponds to $\mathbf{a_0}=\{0,1,0,...,0\}$ and
$Z_1^{(N)}$ corresponds to $\mathbf{a_0}=\{0,0,0,...,1\}$.
As before, the characteristic frequencies are obtained by imposing $\det{\mathbf{S}(\omega_0)}=0$, i.e. Eq.~\eqref{det_modes}.

In a region where the radius of convergence of the series is large enough, the series method is extremely efficient. Indeed, it only requires to compute $N$ Frobenius series in the form~\eqref{Frobenius} for a given truncation order and to construct the complex single-valued function $\det{\mathbf{S}(\omega)}$. Then, a standard shooting method can be implemented to compute the root. In the case of AdS BHs, the radius of convergence strongly depends on the size of the BH with respect to the AdS radius. When $r_+\gg L$, the series converges quickly and the method is efficient. Fortunately, this is the case of major interest for holographic applications. On the other hand, the convergence properties of the Frobenius series are poor for small BHs, and the method becomes practically inefficient when $r_+\ll L$. In the latter case, the spectrum supports slowly-damped modes which can be computed through a Breit-Wigner method~\cite{Berti:2009wx}.
A pedagogical implementation of the matrix-valued series method is presented in the notebook~\url{series_method_DCS.nb}~\cite{webpage}, where we compute the QNMs of a Schwarzschild-AdS BH in DCS gravity [see also next section].


\subsection{Example: QNMs of Schwarzschild BHs in Dynamical Chern-Simons gravity}\label{sec:example}
It is instructive to consider on an example in which the perturbation equations form a coupled system of ODEs.
For concreteness, we focus on a prototype theory, in which the Einstein-Hilbert action is modified by adding an extra scalar field coupled to higher-curvature terms. We consider the following Lagrangian density:
\begin{equation}
 {\cal L}_{\rm DCS}=\sqrt{-g}\left( \frac{R}{16\pi} - \frac{1}{2}\nabla_\rho\phi\nabla^\rho\phi +\frac{\alpha}{4} \phi {}^*RR-V(\phi)\right)+{\cal L}_{\rm matter}\,,\label{actionCS}
\end{equation}
where ${}^*RR\equiv R_{\mu\nu\rho\sigma}{}^*R^{\nu\mu\rho\sigma}=\epsilon_{\sigma\rho\tau\eta}{R_{\mu\nu}}^{\tau\eta}R^{\mu\nu\rho\sigma}/2$. This theory is usually referred to as Dynamical Chern-Simons gravity~\cite{Alexander:2009tp}. Interestingly, it admits all spherically symmetric GR solutions while it deviates from GR in case of rotation. 
In addition, even though spherically symmetric BHs are described by the Schwarzschild metric as in GR, their linear perturbations obey different equations and, in particular, gravitational and scalar perturbations are coupled to each other in this theory.

The Lagrangian density above is an example of what we generically expect in modified theories of gravity in a Lagrangian formulation:
\begin{equation}
 {\cal L}={\cal L}(g_{\mu\nu},\partial_\sigma g_{\mu\nu},...,\phi,\partial_\sigma\phi,...)\,,
\end{equation}
where $\phi$ represents an extra fundamental field of generic spin and, in the case of DCS gravity, it is a pseudoscalar field. While it is straightforward to derive the field equations from the Lagrangian above, the procedure can be lengthy and tedious depending on the extra terms in the Lagrangian. It is then particularly useful to obtain the equations with a symbolic manipulation software. This is presented in the notebook~\url{field_eqs.nb}~\cite{webpage}, which makes use of the external package {\scshape xTensor}~\cite{xTensor} for tensorial calculus\footnote{\textbf{Exercise:} adapt the code to derive the field equations of so-called quadratic gravity~\cite{Yunes:2009hc,Pani:2011gy}, ${\cal L}=\sqrt{-g}(R+\alpha_1 R^2+\alpha_2 R_{\mu\nu}^2+\alpha_3 R_{\mu\nu\rho\sigma}^2+\alpha_4 {}^*RR)$. In the small-$\alpha_i$ limit, what is the differential order are the field equations?}. 

The field equations read~\cite{Cardoso:2009pk,Molina:2010fb}
\begin{eqnarray}
 G_{\mu\nu}&=&8\pi T_{\mu\nu}+8\pi\left[\partial_\mu\phi\partial_\nu\phi-\frac{g_{\mu\nu}}{2}(\partial\phi)^2-g_{\mu\nu}V(\phi)\right]-16\pi\alpha C_{\mu\nu}\,,\\
 \square\phi&=&V'(\phi)-\frac{\alpha}{4}{}^*RR
\end{eqnarray}
where 
\begin{equation}
 C^{\mu\nu}=\nabla_\rho\phi\epsilon^{\rho\sigma\tau(\mu}\nabla_\tau R^{\nu)}_{~~\sigma}+\nabla_\rho\nabla_\sigma\phi\,^*R^{\sigma(\mu\nu)\rho}\,.\label{Ctensor}
\end{equation}
Since in any spherically symmetric background $C^{\mu\nu}\equiv0$ and $\epsilon_{\rho\sigma \tau \eta}{R_{\mu\nu}}^{\tau \eta}R^{\mu\nu\rho\sigma}\equiv0$, spherically symmetric GR solutions are also solutions of this theory. In the following we consider the case $V(\phi)=\Lambda/(8\pi)$ in order to allow for a possible cosmological constant. Therefore, a vanishing scalar field and a Schwarzschild-(A)dS metric,
\begin{equation}
 F(r)=B(r)^{-1}=1-\frac{2M}{r^2}+\frac{\Lambda}{3} r^2\,.\label{backgroundDCS}
\end{equation}
are a consistent vacuum solution of the field equations\footnote{\textbf{Exercise:} They are also the \textit{only} static vacuum solution. Why?}.

Let us apply the harmonic decomposition discussed above to the case of gravito-scalar perturbations of a Schwarzschild BH in DCS gravity. We consider the vacuum case ($T_{\mu\nu}\equiv0$).
Together with metric perturbations, we also decompose the scalar field in spherical harmonics as:
\begin{equation}
 \phi(t,r,\vartheta,\varphi)=\frac{\Theta^\ell(r,t)}{r}Y^{\ell}(\vartheta,\varphi)\,.\label{expscal}
\end{equation}
The perturbation equations are derived in the notebook \url{DCS_pert_eqs.nb}~\cite{webpage}, where we insert the harmonic decomposition of the metric and of the scalar field into the field equations and solve them at first order in the perturbations. In the frequency-domain the gravitational-axial sector and the scalar sector are described by the following system~\cite{Molina:2010fb}
\begin{equation}\label{systemDCS}
 \left\{
 \begin{array}{c}
\left[\frac{d^2}{d r_*^2}+\omega^2-V_{RW}(r)\right]\!Q^{\ell}(t,r)=T_{RW}(r)\Theta^{\ell}(t,r)\\
\left[\frac{d^2}{d r_*^2}+\omega^2-V_{S}(r)\right]\!\Theta^{\ell}(t,r)=T_S(r)Q^{\ell}(t,r)
\end{array}
\right.
\end{equation}
where $r_*$ is the tortoise coordinate defined by $dr/dr_*=F$, $Q^\ell$ is the Regge-Wheeler function in terms of which $h_0^\ell$ and $h_1^\ell$ can be expressed, and the potentials read
\begin{eqnarray}
&&V_{RW}(r)=F\left(\frac{\ell(\ell+1)}{r^2}-\frac{6M}{r^3}\right)\,,\label{potentialRW}\\
&&T_{RW}(r)=F\frac{96i\pi M\omega \alpha}{r^5}\,,\\
&&V_{S}(r)=F\left(\frac{\ell(\ell+1)}{r^2}\left[1+\frac{576\pi
M^2\alpha^2}{r^6\beta}\right]+\frac{2M}{r^3}-\frac{2\Lambda}{3}\right)\,,\label{potentialScalar}\\
&&T_S(r)=-F\frac{(\ell+2)!}{(\ell-2)!}\frac{6Mi\alpha}{r^5\beta\omega}\,. 
\end{eqnarray}
On the other hand, the polar sector is described by the same Zerilli equation as in GR, and it can be solved by standard methods and gives the well-known QNMs of a Schwarzschild BH~\cite{Berti:2009kk}.

The system~\eqref{systemDCS} is already in the form~\eqref{system} and can be solved with the methods described above. In the asymptotically flat case, $\Lambda=0$, the modes can be computed via matrix-valued direct integration (cf.~\url{DCS_DI.nb}). When $\Lambda<0$, the gravito-scalar modes of the Schwarzschild-AdS background can be computed via a matrix version of the series method (cf.~\url{series_method_DCS.nb}). In the corresponding notebooks the dependence of the modes from the coupling constant $\alpha$ is shown.
\section{Perturbations of spinning BHs}	\label{sec:spinning}
In this section we discuss perturbations of \emph{generic} stationary and axisymmetric spacetimes. As discussed in the introduction, this topic is still largely an open problem and, besides the special case of Kerr metric in four-dimensional GR, not much is known on perturbations of other spinning geometries. 

For any stationary and axisymmetric spacetime, the linearized field equations reduce to a set of coupled partial differential equations which depend on the radial coordinate $r$, on the angular coordinate $\vartheta$ and on the time $t$. In the time-domain, one then needs to perform a $2+1$ evolution of the coupled system, whereas in the frequency domain one is left with a two-dimensional boundary problem in the $(r,\vartheta)$ variables. 

Robust methods to evolve $2+1$ coupled systems are available~\cite{Yoshino:2012kn,Strafuss:2004qc,Witek:2012tr} and will be covered elsewhere~\cite{Hiro,Helvi}. Such simulations are numerically challenging and time consuming. Furthermore, with the current computational power it is possible to follow the evolution on a timescale not larger than $10^4M$ with sufficient precision~\cite{Yoshino:2012kn,Strafuss:2004qc,Witek:2012tr}. This is usually sufficient for many purposes, for example for some stability analysis or to extract the characteristic frequencies through a spectral decomposition. However, in other relevant situations one wishes to perform longer and precise simulations, up to $t\sim 10^6 M$ or more. An emblematic example is the evolution of a massive Klein-Gordon field around a Kerr BH~\cite{Dolan:2012yt}. It is well-known that the system develops an exponentially-growing instability which is caused by superradiant amplification of low-frequency waves near the Kerr horizon~\cite{Teukolsky:1974yv,Press:1972zz,Detweiler:1980uk,Cardoso:2004nk,Dolan:2007mj}. The maximum $e$-fold time of the instability is about $10^6M$ which requires a long and stable evolution to capture the exponential growth. Shorter evolutions may lead to misinterpretion of the instability timescale, especially because the time-domain signal shows  beating effects due to interference of several quasi-bound state modes with similar frequencies~\cite{Witek:2012tr}.

The aim of the following sections is to introduce some recent tools that can be viewed as alternative and complementary techniques to ``hard numerics''\footnote{The distinction between ``soft numerics'' and ``hard numerics'' is an interesting concept that emerged during the first NR/HEP Workshop~\cite{Cardoso:2012qm}.}. The latter requires advanced numerical methods that are covered in other work presented at the School~\cite{webpage} and we refer to other lecture's notes~\cite{Hiro,Helvi,Miguel} on this topic.

In Section~\ref{sec:time} we briefly review an analytical technique to transform some $2+1$ problems into simpler $1+1$ problems at \emph{any} order in the spin parameter and the numerical tools required to achieve stable extra-long ($t\sim 10^6M$) evolutions of the field equations. As an example, we shall focus on the case of massive scalar perturbations of a Kerr BH~\cite{Dolan:2012yt}. In more complicated settings, this dimensional reduction might be impractical and other approximation schemes are required.
In Section~\ref{sec:slowrot} we introduce a slow-rotation expansion of the linearized perturbation equations. Considering an arbitrary power of the dimensionless spin parameter, the dynamics of \emph{any} small perturbation propagating on a generic stationary and axisymmetric spacetime can be reduced to a $1+1$ problem in the time-domain, or to a simple one-dimensional eigenvalue problem in the frequency domain.  Finally, in Section~\ref{sec:spectral} we present spectral methods that have been recently applied to the stability analysis of higher-dimensional spinning BHs~\cite{Dias:2009iu,Dias:2010eu}. The variety of analytical and numerical tools we present are in some sense complementary to each other and best-suited for different classes of problems.

\subsection{Reduction to a $1+1$ problem}\label{sec:time}
In this section, we briefly discuss the time evolution of the linearized equations on a stationary and axisymmetric geometries.
It is not our scope to give a detailed presentation, but we simply review some recent developments to perform extra-long evolutions of wavepackets propagating on a spinning BH background~\cite{Dolan:2012yt}. We briefly present the main ideas and refer to other works for a more detailed discussion.

Given a $2+1$ problem where the radial and angular dependences are not manifestly separable, it is nonetheless possible to expand all perturbation variables in a complete basis of spherical harmonics. As we shall discuss in detail, the orthogonality properties of the spherical harmonics can be used to eliminate the angular dependence, at cost of introducing $\ell$-mode-mixing couplings. In principle, this procedure can be performed for any stationary and axisymmetric spacetime, at least when the background is known in closed form. Therefore, the dimensionality of the problem can be lowered to $1+1$ dimensions.

\subsubsection{Example: Massive scalar perturbations of Kerr BHs}
A particularly illuminating example is the case of massive Klein-Gordon probe-fields on a Kerr BH~\cite{Dolan:2012yt}. Following Dolan's original work, here we focus on this case, but the combination of analytical and numerical tools that are presented can find application to a variety of other interesting problems.

We start with the massive Klein-Gordon equation
\begin{equation}
\square\phi=\mu^2\phi\,,\label{KG}
\end{equation}
on the exact Kerr metric in Boyer-Lindquist coordinates:
\begin{eqnarray}
 &&ds_{\rm Kerr}^2=-\left(1-\frac{2Mr}{\Sigma}\right)dt^2
+\frac{\Sigma}{\Delta}dr^2-\frac{4rM^2}{\Sigma}\tilde{a}\sin^2\vartheta d\varphi dt   \nonumber\\
&&+{\Sigma}d\vartheta^2+
\left[(r^2+M^2\tilde{a}^2)\sin^2\vartheta +
\frac{2rM^3}{\Sigma}\tilde{a}^2\sin^4\vartheta \right]d\varphi^2\,,\label{kerrmetric}
\end{eqnarray}
where $\Sigma=r^2+M^2\tilde{a}^2\cos^2\vartheta$, $\Delta=(r-r_+)(r-r_-)$ and $r_\pm=M(1\pm\sqrt{1-\tilde a^2})$. In Eq.~\eqref{KG} $\mu$ is related to the mass $m_s$ of the scalar field by $\mu=m_s \hbar c /G$. In natural units, the quantity $M\mu$ is dimensionless.

In an axisymmetric background, the scalar field can be decomposed as
\begin{equation}
 \phi(t,r,\vartheta,\varphi)=\frac{\Psi(t,r,\vartheta)}{r}e^{im\varphi}\,,
\end{equation}
and the field equation~\eqref{KG} can be written as:~\cite{Dolan:2012yt}
\begin{equation}
 \left[{\cal D}_{tr}-\Delta\left({\cal 
D}_{\vartheta\tilde\varphi}-V-V_c-\mu^2 r^2 \right) \right](\Psi 
e^{i m\tilde\varphi})=0\,,\label{KGtime}
\end{equation}
with
\begin{eqnarray}
{\cal 
D}_{tr}&\equiv&(r^2+\tilde{a}^2M^2)^2[\partial_{tt}
-\partial_ { r_*r_* } ]-M^2\tilde{a}^2\Delta\partial_{tt} +4i\tilde{a} 
mM^2 r\partial_t\nn\\
&&-\left[2i\tilde{a}Mm(r^2+\tilde{a}^2M^2)-\frac{2\tilde{a}^2M^2\Delta}{r}\right]\partial_{r_*}\,,\\
{\cal D}_{\vartheta\tilde\varphi}&\equiv&\partial_{\vartheta\vartheta}+\cot\vartheta\partial_\vartheta+\frac{\partial_{\tilde\varphi\tilde\varphi}}{\sin^2\vartheta}\,,\\
V&\equiv& \frac{2M}{r}\left(1-\frac{\tilde{a}^2M}{r}\right)+\frac{2i\tilde{a}mM}{r}\,,\\
V_c&\equiv& \tilde{a}^2M^2(\mu^2+\partial_{tt})\cos^2\vartheta\,,
\end{eqnarray}
where $r_*$ is the tortoise coordinate defined by $dr/dr_*=\Delta/(r^2+\tilde{a}^2M^2)$ and a new azimuthal coordinate, $d\tilde{\varphi}=d\varphi+\tilde{a}M dr/\Delta$ has been introduced~\cite{Dolan:2012yt}. 
In the frequency domain, the radial and angular dependence of the equation above can be completely separated using a basis of spheroidal harmonics~\cite{Detweiler:1980uk,Dolan:2007mj}. However, the spheroidal eigenvalues are frequency-dependent and this prevents a similar separation in the time domain. Without further reduction, Eq.~\eqref{KGtime} is suitable for standard $2+1$ evolution in the time domain. 

The problem can be made computationally less demanding by 
reducing the dimensionality. This can be achieved by further expanding 
the field in \emph{spherical} harmonics:
\begin{equation}
 \Psi(t,r,\vartheta) e^{im\tilde\varphi}=\psi_\ell(t,r)Y^\ell(\vartheta,\tilde\varphi) \label{KGdec}
\end{equation}
(hereafter a sum over $m$ and $\ell\geq|m|$ is understood) and noting 
that the spherical harmonics $Y^\ell$ are eigenfunctions of ${\cal 
D}_{\vartheta\tilde\varphi}$, which is the Laplace operator on the 
sphere [cf. Eq.~\eqref{propY} in~\ref{app:orthogonality}]. If 
$V_c\equiv0$, this would be sufficient to separate the angular part 
and obtain a single radial equation for each index $\ell$.

However, the term $\cos^2\vartheta$ in $V_c$ prevents this decoupling. Nonetheless, such term can be written as a 
combination of spherical harmonics with harmonic indices $\ell$ and 
$\ell\pm2$. To show this, we start by the following property of the 
scalar spherical harmonics:
\begin{equation}
\cos\th Y^{\ell}=\cQ_{\ell+1}Y^{\ell+1}+\cQ_{\ell}Y^{\ell-1}\,,\label{ident1}
\end{equation}
where 
\begin{equation}
{\cal Q}_\ell=\sqrt{\frac{\ell^2-m^2}{4\ell^2-1}}\,.\label{Qpm}
\end{equation}
By repeated use of the equation above, we obtain\footnote{\textbf{Exercise:} derive a similar relation for $\cos^n\th Y^{\ell}$.}
\begin{equation}
\cos^2\th Y^{\ell}=\left(\cQ_\lp^2 + \cQ_\ell^2\right)Y^\ell+\cQ_\lp \cQ_\lpp Y^\lpp + \cQ_\ell \cQ_\lm Y^\lmm\,.\label{ident3}
\end{equation}
Other similar identities are listed in Eqs.~\eqref{ident2}--\eqref{ident4} below. Using the property above, Eq.~\eqref{KGtime} can be written as
\begin{eqnarray}
 &&Y^\ell\left[\frac{{\cal D}_{tr}}{\Delta}-\ell(\ell+1)-V-\mu^2 r^2\right]\psi_\ell\nn\\
 &&=\tilde{a}^2M^2(\mu^2+\partial_{tt})\left[\left(\cQ_\lp^2 + 
\cQ_\ell^2\right)Y^\ell+\cQ_\lp \cQ_\lpp Y^\lpp + \cQ_\ell \cQ_\lm 
Y^\lmm\right]\psi_\ell\,.
\end{eqnarray}
Finally, using the orthogonality properties of the spherical harmonics,
\begin{equation}
 \int Y^{\ell} Y^{*\,\ell'}d\Omega=\delta^{\ell\ell'}\,,\label{orthoscalar}
\end{equation}
and integrating over the two-sphere [cf. also~\ref{app:orthogonality}], we obtain the final equation:
\begin{eqnarray}
 &&{\cal D}_{tr}\psi_\ell-\Delta \left\{\ell(\ell+1)+V+\mu^2 r^2 
+\tilde{a}^2M^2\left(\cQ_\lp^2 + 
\cQ_\ell^2\right)(\mu^2+\partial_{tt})\right\}\psi_\ell\nn\\
 &&=\tilde{a}^2M^2\Delta(\mu^2+\partial_{tt})\left[\cQ_\ell 
\cQ_{\ell-1}\psi_{\ell-2}  + \cQ_{\ell+2} \cQ_{\ell+1} 
\psi_{\ell-2}\right]\,. \label{KGfinaltime}
\end{eqnarray}
The angular dependence has been completely eliminated and the problem has been reduced to a $1+1$ equation. However, in this equation the field $\psi_\ell$ is coupled to $\psi_{\ell\pm2}$. Because $\ell\geq|m|$, Eq.~\eqref{KGfinaltime} actually contains an \emph{infinite} number of coupled equations. In practice, the coupled system can 
be solved by truncating the sum over $\ell$ implicit in Eq.~\eqref{KGdec} to some order $L$, and checking convergence of the results when $L$ is sufficiently large~\cite{Pani:2012vp,Dolan:2012yt}.

\subsubsection{Time evolution}
A standard approach to solve equations in the form of Eq.~\eqref{KGfinaltime} is to use the so-called method of lines and a finite-difference approximation on spatial slides~\cite{NumericalRecipes}. Defining a one-dimensional grid along the radial direction, spatial derivatives are substituted with finite differences of various order. The system is then reduced to a second-order-in-time set of ODEs. The system can be reduced to first-order form:
\begin{equation}
 \frac{d\mathbf{y}(t)}{dt}=\mathbf{A}\mathbf{y}(t)\,, \label{MoL}
\end{equation}
where $\mathbf{y}$ is a vector containing the variables $\psi_\ell$ and their momenta $\partial_t \psi_\ell$ discretized on the grid. We refer to classical books~\cite{NumericalRecipes} and to other lecture notes~\cite{Helvi,Hiro} for advanced methods to solve this class of problems. 
\subsubsection{Spectral analysis}
Assuming a stable time evolution of the system~\eqref{MoL} is achieved, it is useful to perform a Fourier analysis of the waveform in order to extract the eigenfrequency spectrum and the amplitudes of the single modes.
The power spectrum at a given frequency and for a given harmonic index $\ell$ is
\begin{equation}
 P_\ell(\omega)=|f_\ell(\omega)|^2\,,\label{power}
\end{equation}
where the Fourier amplitude reads
\begin{equation}
 f_\ell(\omega_j)=\frac{1}{N}\sum_{k=0}^{N-1}\psi_\ell(t_k) e^{-i\omega_j t_k}\,,\label{fourier}
\end{equation}
where $\psi_\ell$ is evaluated at a fixed radial position, $t_k=k \Delta t$ with $\Delta t=t_F/(N-1)$ and we assume a time evolution in the domain $[0,t_F]$ discretized in $N$ equidistant points~\cite{Dolan:2012yt}. The resolution in frequency is given by $2\pi/t_F$, so that the longer the simulation the more refined is the frequency spectrum.

Near an eigenfrequency $\omega_R+i\omega_I$, the power spectrum --~considered as a function on the real axis~-- has the typical Breit-Wigner form:
\begin{equation}
 P(\omega)\approx \frac{1}{(\omega-\omega_R)^2+\omega_I^2}\,,\qquad \omega\approx \omega_R \label{PowerBW}
\end{equation}
and the real and imaginary parts can be extracted through a quadratic fit around the power peak. This procedure is more precise when $\omega_I\ll\omega_R$, so that $\omega_R$ is approximately a pole of the power spectrum and the width of the resonance is related to the imaginary part.
To achieve more precision, filtering techniques may be used. A simple example is explained in Ref.~\cite{Dolan:2012yt}, where the modes in the Fourier amplitude are isolated using a filter peaked at $\omega_R$. An inverse Fourier-transform is performed on the filtered signal, and the imaginary part can be precisely extracted from the waveform at intermediate time. The results obtained by this technique are quite impressive, as they allow to extract the fundamental unstable mode and the first overtones with good precision, even when the evolution is followed only up to one tenth of the instability timescale~\cite{Dolan:2012yt}.
\subsubsection{Limitations}
The $1+1$ reduction discussed above can be applied to other linearized equations and to other metric backgrounds, at least as long as the angular dependence of the perturbation equations can be written in terms of simple trigonometric functions. However, if the field equations are more involved (for example in the case of massive spin-1 fields~\cite{Rosa:2011my,Pani:2012vp,Pani:2012bp} or in more general cases) such harmonic decomposition would introduce couplings not only between the nearest-neighbor modes, but also to the next-to-nearest ones and so on, and it would also introduce parity-mixing couplings. In this case a higher truncation order $L$ might be needed and this would result in a large number of $1+1$ coupled differential equations, which may be challenging to evolve for long times. Furthermore, the convergence properties of the solution at a given truncation order might deteriorate. Nonetheless, in some cases this procedure can still be more convenient (or complementary) to a brute-force $2+1$ evolution. 

An alternative method to reduce more involved problems is to introduce a series of infinite couplings to higher $\ell$ modes~\cite{Racz:2011qu}. If the coupling terms die away sufficiently fast, the coupled system is still suitable for stable evolution.

To overcome these difficulties, in the next section we introduce a further approximation scheme and we consider a slow-rotation expansion of the linearized equations. As we shall discuss, the slowly-rotating framework simplifies the perturbation equations considerably and it is well suited to attack arbitrarily complicated systems of coupled equations.
%

\subsection{Slow-rotation expansion}\label{sec:slowrot}
In this section we discuss a general method to study linear perturbations of slowly
rotating BHs that is particularly useful when the perturbation
variables are not separable. The method is an extension of Kojima's work on
perturbations of slowly rotating neutron stars
\cite{Kojima:1992ie,1993ApJ...414..247K,1993PThPh..90..977K} and it has been recently generalized and put on firmer basis in the context of BH perturbations~\cite{Pani:2012vp,Pani:2012bp}. The idea is that slowly-rotating backgrounds are ``close enough'' to spherical symmetry that an approximate separation of the perturbation equations in radial and angular parts becomes possible. Similarly to the case discussed in the previous section, the perturbation functions are expanded in spherical harmonics and they reduce, in general, to a $1+1$ coupled system of differential equations where various couplings between different multipolar indices $\ell$ and between perturbations with different parity are introduced. 

However, at variance with what discussed above, the slow-rotation approximation guarantees that only a certain (typically small) number of couplings to higher multipoles contributes to a given order in $\tilde{a}\ll1$. This makes the method well suited to investigate complicated systems of coupled equations. In the Fourier space, one is left with a simple system of ODEs which can be integrated by standard methods.

\subsubsection{Formalism}
Let us start by considering the most general stationary axisymmetric spacetime~\cite{Chandra}
\begin{equation}
 ds^2_0=-H^2dt^2+Q^2dr^2+r^2K^2\left[d\vartheta^2+\sin^2\vartheta(d\varphi-Ldt)^2\right]\,,\nn
\end{equation}
where $H$, $Q$, $K$ and $L$ are functions of $r$ and $\vartheta$
only. 
This metric can describe a stationary, axisymmetric compact object (such as a neutron star or a BH).
If the object is slowly rotating, one can define a perturbative expansion in the angular momentum $J$ (or in some other parameter linear in $J$, which characterizes the rotation rate).

To second order in rotation, the metric above can be expanded
as~\cite{Hartle:1967he}
\begin{align}
ds^2_0=&-F(r)\left[1+F_2\right]dt^2 +B(r)^{-1}\left[1+\frac{2B_2}{r-2M}\right]dr^2\nn\\
&+r^2(1+k_2)\left[d{\vartheta}^2+\sin^2\th(d\varphi-\varpi dt)^2\right]	\,,
\label{metric2}
\end{align}
where $M$ is the mass of the spacetime, $\varpi$ is a function of $r$
linear in the rotation parameter, and $F_2$, $B_2$ and $k_2$ are
functions of $r$ and $\th$ quadratic in the rotation parameter.
The functions $H$, $K$ and $Q$ transform like scalars under rotation, and
they can be expanded in scalar spherical harmonics which, due to
axisymmetry, reduce to the Legendre polynomials $P_\ell(\th)$. To second order, only $\ell=0$ and $\ell=2$
polynomials contribute~\cite{Hartle:1967he}, therefore we obtain:
\begin{eqnarray}
 F_2(r,\vartheta)&=&F_{r}(r)+F_{\vartheta}(r)P_2(\vartheta)\,,\\
  B_2(r,\vartheta)&=&B_{r}(r)+B_{\vartheta}(r)P_2(\vartheta)\,,\\
   k_2(r,\vartheta)&=&k_{r}(r)+k_{\vartheta}(r)P_2(\vartheta)\,.
\end{eqnarray}
At first order in rotation the metric~\eqref{metric2} reduces to a much simpler form
\begin{align}
ds^2_0=-F(r)dt^2 +B(r)^{-1}dr^2-2\varpi(r)\sin^2\th d\varphi dt+r^2d^2\Omega\,.\label{metric1}
\end{align}
The metric~\eqref{metric2} and~\eqref{metric1} can be computed solving Einstein's equations to second and first order in the rotation, respectively. For example, to first order, the slowly-rotating Kerr metric corresponds to
\begin{equation}
 F(r)=B(r)=1-2M/r\,,\quad \varpi=2M^2\tilde a/r \,, \label{Kerr1st}
\end{equation}
where $M$ and $J=\tilde a M^2$ are the mass and the angular momentum of the BH.
More generically, given a nonrotating metric the gyromagnetic function $\varpi(r)$ can be computed using the approach originally developed by
Hartle~\cite{Hartle:1967he}. This is the case for some BH solutions in modified gravity theories, which are only known perturbatively~\cite{Yunes:2009hc, Pani:2009wy,Pani:2011gy,Yagi:2012ya}. The metric can also be constructed numerically, for instance in the case of slowly-rotating stars~\cite{Kojima:1992ie,1993ApJ...414..247K,1993PThPh..90..977K}.

Slowly rotating and oscillating compact objects can be studied as perturbations of the axisymmetric, stationary solutions discussed above.
Scalar, vector and tensor field equations in the background
metric~\eqref{metric2} can be linearized in the field perturbations.
Any perturbation function $\delta X$ can be expanded in a complete basis of spherical harmonics, as previously discussed for the nonrotating case. Schematically, in the frequency domain we have
\begin{equation}
\delta X_{\mu_1\dots}(t,r,\vartheta,\varphi)=
\delta X^{(i)}_{\ell m}(r){\cal Y}_{\mu_1\dots}^{\ell m\,(i)}e^{-i\omega t}\,,
\label{expa}
\end{equation}
where ${\cal Y}_{\mu_1\dots}^{\ell m\,(i)}$ is a basis of scalar,
vector or tensor harmonics, depending on the tensorial nature of the
perturbation $\delta X$. As in the spherically symmetric case, the perturbation variables $\delta X^{(i)}_{\ell m}(r)$ can
be classified as ``polar'' or ``axial'' depending on their behavior
under parity transformations.

The linear response of the system is fully characterized by a coupled system of ODEs in the perturbation functions $\delta X^{(i)}_{\ell  m}(r)$.  
As previously discussed, in the case of a spherically symmetric background,
perturbations with different values of $(\ell,\,m)$, as well as
perturbations with opposite parity, are decoupled. In a rotating,
axially symmetric background, perturbations with different values of
$m$ are still decoupled but perturbations with different values of
$\ell$ are not.

At this stage, we present the general schematic form of the perturbation equations, and postpone the derivation of some particular cases to Section~\ref{sec:applications}. To second order, the perturbation equations schematically read
\begin{eqnarray}
0&=&{\cal A}_{\ell}+\tilde a m \bar{\cal A}_{{\ell}}+\tilde{a}^2 \hat{{\cal A}}_\ell\nn\\
&+&\tilde a ({\cal Q}_{{\ell}}\tilde{\cal P}_{\ell-1}+{\cal Q}_{\ell+1}\tilde{\cal P}_{\ell+1})\nn\\
&+&\tilde{a}^2 \left[\cQ_\lm \cQ_\ell \breve{{\cal A}}_\lmm + \cQ_\lpp \cQ_\lp 
\breve{{\cal A}}_\lpp \right]+{\cal O}(\tilde{a}^3)\,,\nn\\\label{epF1c}\\
0&=&{\cal P}_{\ell}+\tilde a m \bar{\cal P}_{{\ell}}+\tilde{a}^2 \hat{{\cal P}}_\ell\nn\\
&+&\tilde a ({\cal Q}_{{\ell}}\tilde{\cal A}_{\ell-1}+{\cal Q}_{\ell+1}\tilde{\cal A}_{\ell+1})\nn\\
&+&\tilde{a}^2 \left[\cQ_\lm \cQ_\ell \breve{{\cal P}}_\lmm + \cQ_\lpp \cQ_\lp 
\breve{{\cal P}}_\lpp \right]+{\cal O}(\tilde{a}^3)\,,\nn\\\label{epF2c}
\end{eqnarray}
where ${\cal Q}_\ell$ were defined in Eq.~\eqref{Qpm}, and ${\cal A}_{\ell}$, $\bar {\cal A}_{\ell}$, $\tilde {\cal A}_{\ell}$,
$\hat {\cal A}_{\ell}$, $\breve {\cal A}_{\ell}$ are \emph{linear}
combinations of the axial perturbations with
multipolar index $\ell$; similarly, ${\cal P}_{\ell}$, $\bar {\cal
  P}_{\ell}$, $\tilde {\cal P}_{\ell}$, $\hat {\cal P}_{\ell}$,
$\breve {\cal P}_{\ell}$ are linear combinations of the polar
perturbations with index $\ell$. 

The structure of Eqs.~\eqref{epF1c}--\eqref{epF2c} is very interesting.
In the limit of slow rotation there is a Laporte-like
``selection rule''~\cite{ChandraFerrari91}: at first order in
$\tilde{a}$, perturbations with a given value of $\ell$ are only
coupled to those with $\ell\pm1$ and \emph{opposite} parity. At second order, perturbations with a
given value of $\ell$ are also coupled to those with $\ell\pm2$ and
\emph{same} parity, and so on. More precisely, perturbations with a given parity and index $\ell$ are coupled to: (i)
perturbations with \emph{opposite} parity and index $\ell\pm1$ at
order $\tilde{a}$; (ii) perturbations with \emph{same} parity and
\emph{same} index $\ell$ up to order $\tilde{a}^2$; (iii)
perturbations with \emph{same} parity and index $\ell\pm2$ at order
$\tilde{a}^2$. The symmetries of the harmonic expansion guarantee that this scheme is preserved at any order in $\tilde{a}$.

Furthermore, from Eq.~\eqref{Qpm} it follows that ${\cal Q}_{\pm m}=0$, and
therefore if $|m|=\ell$ the coupling of perturbations with index
$\ell$ to perturbations with indices $\ell-1$ and $\ell-2$ is
suppressed. This general property is usually called~\cite{ChandraFerrari91} ``propensity
rule'' in atomic theory, and states that transitions $\ell\to\ell+1$
are strongly favored over transitions $\ell\to\ell-1$. Indeed, the slow-rotation technique is well known in quantum mechanics and the coefficients ${\cal Q}_{\ell}$ are related to the
usual Clebsch-Gordan coefficients.

\subsubsection{Eigenvalue spectrum in the slow-rotation limit}
Due to the coupling between different multipolar indices, the spectrum of the solutions of Eqs.~\eqref{epF1c}--\eqref{epF2c} is extremely rich. However, if we are interested in the characteristic modes of the slowly-rotating background to first or to second order in $\tilde{a}$, the perturbation equations can be considerably simplified.

Let us start by considering the first order corrections. We expand all quantities to first order and we ignore
the terms $\hat {\cal A}_{\ell}$, $\breve {\cal A}_{\ell}$, $\hat
{\cal P}_{\ell}$ and $\breve {\cal P}_{\ell}$, which are multiplied by $\tilde{a}^2$ in Eqs.~\eqref{epF1c} and \eqref{epF2c}.  Crucially, the terms ($\tilde {\cal P}_{\ell},\,\tilde {\cal A}_{\ell}$) do not contribute to the eigenfrequencies at first order
in $\tilde{a}$~\cite{1993PThPh..90..977K,Pani:2012bp}.  
Here, we follow the proof that was presented in Ref.~\cite{Pani:2012bp}.
At first order, Eqs.~\eqref{epF1c} and \eqref{epF2c} are invariant under the simultaneous transformations
\begin{subequations}
\begin{eqnarray}
& a_{\ell m}\to\mp a_{\ell -m}\,,\qquad  p_{\ell m}\to \pm p_{\ell -m}\,,\\
& \tilde{a}\to-\tilde{a}\,,\qquad \hspace{1cm} m\to -m\,,
\end{eqnarray}
\end{subequations}
where $a_{\ell m}$ ($p_{\ell m}$) schematically denotes any
axial (polar) perturbation variables with indices $(\ell,m)$. The
invariance follows from the linearity of the terms in
Eqs.~\eqref{epF1c} and \eqref{epF2c} and from the fact that the ${\cal  Q}_{\ell}$'s are \emph{even} functions of $m$. The boundary
conditions that define the characteristic modes of the BH are also
invariant under the transformations above. Therefore in the slow-rotation limit the
eigenfrequencies can be expanded as
%
\begin{equation}
 \omega=\omega_0+m\,\omega_1 \tilde{a}+\omega_2 
\tilde{a}^2+{\cal O}(\tilde{a}^3)\,, \label{exp_omega}
\end{equation}
where $\omega_0$ is the eigenfrequency of the nonrotating spacetime
and $\omega_n$ is the $n$-th order correction\footnote{$\omega_1$
and $\omega_2$ are generically polynomials in $m$ but, due to the
above symmetry, $\omega_1$ is an \emph{even} polynomial.}.
Crucially, only the terms ($\bar {\cal P}_{\ell},\,\bar {\cal
  A}_{\ell}$) in Eqs.~\eqref{epF1c} and \eqref{epF2c} can contribute
to $\omega_1$. Indeed, due to the factor $\tilde{a}$ in front of all
terms ($\bar {\cal P}_{\ell},\,\bar {\cal A}_{\ell}$, $\tilde {\cal
  P}_{\ell},\,\tilde {\cal A}_{\ell}$) and to their linearity, at
first order in $\tilde{a}$ we can simply take the zeroth order (in
rotation) expansion of these terms. That is, to our level of
approximation the terms ($\bar {\cal P}_{\ell},\,\bar {\cal
  A}_{\ell}$, $\tilde {\cal P}_{\ell},\,\tilde {\cal A}_{\ell}$) in
Eqs.~\eqref{epF1c} and \eqref{epF2c} only contain the perturbations of the \emph{nonrotating}, spherically symmetric background. Since the latter do not explicitly depend on $m$, the $m$ dependence in
Eq.~\eqref{exp_omega} can only arise from the terms ($\bar {\cal P}_{\ell},\,\bar {\cal A}_{\ell}$) to zeroth order.

Therefore, the eigenvalue problem to first order is equivalent to the following \emph{decoupled} sets of equations:
\begin{eqnarray}
{\cal A}_{\ell}+\tilde a m \bar{\cal A}_{{\ell}}&=&0\,,\label{epF1}\\
{\cal P}_{\ell}+\tilde a m \bar{\cal P}_{{\ell}}&=&0\,.\label{epF2}
\end{eqnarray}
In the equations above polar and axial perturbations --~as well as
perturbations with different values of the harmonic indices~-- are
decoupled from each other and can be studied independently. In practice, the final eigenvalue problem is very similar to the nonspinning case, the only difference being the introduction of the Zeeman-splitting term proportional to $\tilde{a}m$ that breaks the azimuthal degeneracy.

Let us now move to the the eigenfrequency spectrum at second
order in $\tilde{a}$. First, we expand any axial and polar perturbation function (respectively denoted as $a_{\ell m}$ and $p_{\ell m}$) that appears in Eqs.~\eqref{epF1c} and \eqref{epF2c}:
\begin{eqnarray}
a_{\ell m}&=&a^{(0)}_{\ell m}+\tilde a\,a^{(1)}_{\ell m}+\tilde a^2a^{(2)}_{\ell m}+{\cal O}(\tilde{a}^3)\nn\\
p_{\ell m}&=&p^{(0)}_{\ell m}+\tilde a\,p^{(1)}_{\ell m}+\tilde a^2p^{(2)}_{\ell m}+{\cal O}(\tilde{a}^3)\,.
\end{eqnarray}
The terms $\breve{{\cal A}}_{\ell\pm2}$ and $\breve{{\cal
    P}}_{\ell\pm2}$ are multiplied by factors $\tilde{a}^2$, so they
only depend on the zeroth-order perturbation functions, $a^{(0)}_{\ell
  m}$, $p^{(0)}_{\ell m}$. The terms $\tilde{{\cal A}}_{\ell\pm1}$ and
$\tilde{{\cal P}}_{\ell\pm1}$ are multiplied by factors $\tilde{a}$,
so they only depend on zeroth- and first-order perturbation functions
$a^{(0)}_{\ell m}$, $p^{(0)}_{\ell m}$, $a^{(1)}_{\ell m}$,
$p^{(1)}_{\ell m}$.

Since in the nonrotating limit axial and polar perturbations are
decoupled, a possible consistent set of solutions of the system
(\ref{epF1c})--(\ref{epF2c}) has $a^{(0)}_{\ell\pm2 m}\equiv0$, which leads to the ``axial-led''~\cite{Lockitch:1998nq} subset of Eqs.~(\ref{epF1c})--(\ref{epF2c}):
\begin{equation}\label{epF1bis}
 \left\{ \begin{array}{l}
          {\cal A}_{\ell}+\tilde a m \bar{\cal A}_{{\ell}}+\tilde{a}^2 \hat{{\cal  A}}_\ell+\tilde a ({\cal Q}_{{\ell}}\tilde{\cal P}_{\ell-1}+
{\cal Q}_{\ell+1}\tilde{\cal P}_{\ell+1})=0\,,\\
{\cal P}_{\ell+1}+\tilde a m \bar{\cal P}_{{\ell+1}}+
\tilde a {\cal Q}_{{\ell+1}}\tilde{\cal A}_{\ell}=0\,,\\
{\cal P}_{\ell-1}+\tilde a m \bar{\cal P}_{{\ell-1}}+
\tilde a {\cal Q}_{{\ell}}\tilde{\cal A}_{\ell}=0\,.
         \end{array}\right.
\end{equation}
Similarly, another consistent set of solutions of the same system has
$p^{(0)}_{\ell\pm2 m}\equiv0$. The corresponding ``polar-led'' system reads 
\begin{equation}\label{epF2bis}
 \left\{ \begin{array}{l}
          {\cal P}_{\ell}+\tilde a m \bar{\cal P}_{{\ell}}+\tilde{a}^2 \hat{{\cal P}}_\ell+
\tilde a ({\cal Q}_{{\ell}}\tilde{\cal A}_{\ell-1}+{\cal Q}_{\ell+1}
\tilde{\cal A}_{\ell+1})=0\,,\\
{\cal A}_{\ell+1}+\tilde a m \bar{\cal A}_{{\ell+1}}
+\tilde a {\cal Q}_{{\ell+1}}\tilde{\cal P}_{\ell}=0\,,\\
{\cal A}_{\ell-1}+\tilde a m \bar{\cal A}_{{\ell-1}}+
\tilde a {\cal Q}_{{\ell}}\tilde{\cal P}_{\ell}=0\,,
         \end{array}\right.
\end{equation}
In the second and third equations of the two systems above we have
dropped the $\tilde{{\cal A}}_{\ell\pm2}$ and $\tilde{{\cal P}}_{\ell\pm2}$ terms, because they only enter at zeroth order, and we have set $a^{(0)}_{\ell\pm2 m}\equiv0$ and $p^{(0)}_{\ell\pm2 m}\equiv0$\footnote{Even though in principle there may be modes which do not belong to the classes of ``axial-led'' or ``polar-led'' perturbations, all solutions belonging to one of these classes which fulfill the appropriate boundary conditions defining
QNMs or bound states are also solutions of the full system
\eqref{epF1c}--\eqref{epF2c} and belong to the eigenspectrum (up to second order in $\tilde{a}$).
}.
Interestingly, within this perturbative scheme a notion of ``conserved quantum number'' $\ell$ is still meaningful, even though, for any given $\ell$, rotation couples terms with opposite parity and different multipolar index. 

To summarize, the eigenfrequencies (or at least a subset of the
eigenfrequencies) of the general system~\eqref{epF1c}, \eqref{epF2c}
can be found, at first order in $\tilde{a}$, by solving the two
decoupled sets~\eqref{epF1} and~\eqref{epF2} for axial and polar
perturbations, respectively. At second order in $\tilde{a}$ we must
solve either the set~\eqref{epF1bis} or the set
~\eqref{epF2bis} for ``axial-led'' and
``polar-led'' modes, respectively.  The power of this procedure stands in its generality. It can be applied to any slowly-rotating spacetime
and to any kind of perturbation. In Section~\ref{sec:applications}, we explicitly derive some particular cases.

\subsection{Stability of higher-dimensional BHs: Spectral methods}\label{sec:spectral}
Although a detailed overview is beyond the scope of this work, we wish to conclude this section by mentioning another method that was recently developed to study the instability of highly-spinning BHs in higher dimensions~\cite{Dias:2009iu,Dias:2010eu}. In $D>4$, thermodynamical and perturbative arguments suggest that quasi-extremal or highly-spinning BH geometries should be linearly unstable~\cite{Emparan:2008eg}. A slow-rotation approximation is not promising to study the instability if the latter is a high-spin effect. Furthermore, the parameter space of the perturbations in higher dimensions can be extremely large and a complete characterization of the linear dynamics is not feasible. Finally, as previously discussed the angular part of the perturbation equations does not appear to be separable in the general case, which makes the linearized problem particularly challenging.

In the approaches taken so far, one restricts the analysis to some special subclass of perturbations that preserve some rotational symmetry of the background. In the frequency domain, the idea is to reduce the linearized dynamics to a two-dimensional boundary problem involving coupled partial differential equations or coupled ODEs, depending on the background geometry. Such problem can be efficiently solved by a Chebyshev spectral method~\cite{Monteiro:2009ke}.

In a nutshell, the method proceeds as follows. First, the spinning BH metric (for instance a Myers-Perry BH with a single spin~\cite{Dias:2009iu} or a cohomogeneity-one Myers-Perry geometry~\cite{Dias:2010eu}) is embedded into a black string with one extra spatial dimension $z$. A subclass of the stationary perturbations of the black string is considered in the form $\sim e^{-i\omega t}e^{i k z}h_{\mu\nu}$, where $h_{\mu\nu}$ does not depend on $t$ and $z$. In the transverse and traceless gauge, the linearized Einstein equations have the form~\cite{Dias:2009iu,Dias:2010eu}
\begin{equation}
 (\Delta_L h)_{\mu\nu}\equiv -\nabla_\rho\nabla^\rho h_{\mu\nu}-2R_{\mu\nu\rho\sigma} h^{\rho\sigma}=-k^2 h_{\mu\nu}\,,
\end{equation}
where $\Delta_L$ is the Lichnerowicz operator on the corresponding Myers-Perry background. Perturbations of the original spinning BH metric are obtained when $k=0$.
As usual, after suitable boundary conditions are imposed at the horizon and at infinity, the equations above define an eigenvalue problem for the complex frequency $\omega$. Depending on the background metric, the problem is reduced to a set of coupled partial differential equations~\cite{Dias:2009iu} or a system of coupled ODEs~\cite{Dias:2010eu}. In the latter case the eigenvalue problem can be solved with the methods described in the previous sections. Nonetheless, if one is interested in finding \emph{purely imaginary} modes, $\omega=i\omega_I$, the task can be simplified by reversing the eigenvalue problem. For a given $\omega_I$ and given spin parameter(s) one seeks for the (real) value of $k$ that solves the boundary problem. Modes with $\omega_I>0$ and $k\neq0$ correspond to black-string metrics that are unstable under the Gregory-Laflamme instability~\cite{Emparan:2008eg}. If such modes exist, by increasing the angular momentum it is possible to track the eigenvalues $k$. The critical value of the spin corresponding to $k=0$ (if it exists) signals the onset of an instability of the associated Myers-Perry background. In order to require the existence of purely imaginary modes in first instance, a particular subclass of perturbations must be considered.~\cite{Dias:2009iu,Dias:2010eu}

This method is clearly not optimally suited to explore the whole parameter space, but it can be efficiently adopted to prove the existence of a subclass of unstable modes and to construct marginally stable solutions at the bifurcation point (which are new BH solutions with pinched horizons). Adopting this method, singly-spinning Myers-Perry BHs in $D\geq6$~\cite{Dias:2009iu,Dias:2010maa}, Myers-Perry BHs in $D=9$ with equal angular momenta~\cite{Dias:2010eu} and in $D=7$ with two of the three angular momenta set to be equal~\cite{Dias:2011jg}, were all found to be unstable above a critical value of the spin. In fact, arguments have been provided for a generic ultraspinning instability of Myers-Perry BHs in $D>5$. 
 
Let us briefly discuss the spectral method that can be adopted to solve the boundary value problem~\cite{Monteiro:2009ke,Dias:2010eu,SpectralMethods}. In order to illustrate how the method works, let us consider a generalized boundary problem defined by $N$ coupled ODEs:
\begin{equation}
 \boldsymbol{{\cal D}}\mathbf{Y}=-k^2\mathbf{V}\mathbf{Y}\,,\label{boundarySM}
\end{equation}
where $\boldsymbol{{\cal D}}$ is a $N\times N$ differential operator, $\mathbf{V}$ is a $N\times N$ matrix and $\mathbf{Y}$ is the $N$-dimensional eigenfunction that we want to compute in a finite domain $y\in[y_i,y_f]$. The basic idea of spectral methods is to approximate the eigenfunction by a finite sum of polynomials:
\begin{equation}
 Y_i(y)=\sum_{j=0}^n a_j^{(i)} y^j\,,
\end{equation}
where $Y_i$ is the $i$th component of $\mathbf{Y}$ ($i=1,...,N$). In practice, the coefficients $a_j^{(i)}$ are obtained by a polynomial interpolation. In order for the method to be accurate and stable, it is crucial to interpolate the functions in a suitably-chosen set of points. The best repartition depends on the problem at hand, here we consider the so-called Chebyshev nodes, which are the roots of the Chebyshev polynomials of the first kind~\cite{Monteiro:2009ke,Dias:2010eu}:
\begin{equation}
 y_l=\frac{y_f+y_i}{2}+\frac{y_f-y_i}{2}\cos\left(\frac{(2l-1)\pi}{2n}\right)\,, \label{Chebyshev}
\end{equation}
where $l=0,...,n$. This repartition is particularly useful because the corresponding polynomial interpolation minimizes the Runge's phenomenon~\cite{NumericalRecipes} i.e., roughly speaking, they minimize the errors of higher-order interpolations at the boundary. The interpolation effectively maps $Y_i(y)\to a_j^{(i)}$, where the latter are just constants. Equation~\eqref{boundarySM} can be transformed into an \emph{algebraic} system for the $(n+1)$-dimensional vectors containing $Y_i(y_l)$ as entries and where differential operators are transformed to nondiagonal matrices that mix the various components of the vectors~\cite{SpectralMethods}. Once boundary conditions are imposed on $Y_i(y_0)$ and on $Y_i(y_N)$, i.e. on the first and on the last entries of each vector, the problem effectively reduces to an algebraic eigenvalue problem for $k$ in $N(n-1)$ dimensions~\cite{Dias:2010eu}., which can be solved by standard methods~\cite{NumericalRecipes,SpectralMethods}.

As mentioned above, this method is efficient to find purely imaginary unstable modes, but it is not well-suited to explore the full parameter space. This is because one seeks for purely real values of $k$, and this would require a multivariate search in the complex $\omega$-plane. It would be very interesting to complement this method with a slowly-rotating approximation of the Myers-Perry family. Even though the results of the slow-rotation expansion cannot be extrapolated to the ultra-spinning regime, this method would allow to treat perturbations of higher-dimensional BHs generically and to explore the full parameter space at a given order in $\tilde{a}$. Furthermore, it is not impossible that ultra-spinning instabilities corresponding to purely imaginary modes will have a counterpart in complex unstable modes at lower rotation rate. Possibly, such instabilities can be captured by a slow-rotation analysis. Another interesting application of this approach could be to provide semianalytical arguments in favor of the so-called bar-mode instability of higher-dimensional BHs~\cite{Shibata:2010wz}. 

\section{Applications of the slow-rotation formalism}\label{sec:applications}
In order to illustrate how the slow-rotation framework presented in Section~\ref{sec:slowrot} works, in this
section we work out some simple application.
\subsection{Massive scalar perturbations of slowly rotating Kerr BHs}\label{sec:scalar}
Let us again consider the massive Klein-Gordon equation~\eqref{KG} around a Kerr BH in the frequency domain. 
Even though this problem is separable in the standard Teukolsky approach, one may apply the slow-rotation formalism to this case and check the errors introduced by the slow-rotation approximation by comparing with exact results. Thus, our goal is to derive the perturbation equations up to second order in rotation. The entire derivation reported in this section is also available in the selfconsistent {\scshape Mathematica} notebook \url{slow_rot_scalar.nb}\cite{webpage}.

The exact Kerr metric in Boyer-Lindquist coordinates reads as in Eq.~\eqref{kerrmetric}.
To second order in $\tilde{a}$, the metric can be written in the form~\eqref{metric2} and the event
horizon $r_+$, the Cauchy horizon $r_-$ and the outer ergosphere $r_{S^+}$ respectively read:
\begin{equation}
 r_+=2M\left(1-\frac{\tilde{a}^2}{4}\right)\,,\quad r_-=\frac{M\tilde{a}^2}{2}\,, \quad r_{S^+}=2M\left(1-\cos^2\th\frac{\tilde{a}^2}{4}\right)\,.
\label{cauchyhor}
\end{equation}
Corrections to the horizon location are of second-order and up to first order the ergosphere coincides with the horizon.
Again, we decompose the scalar field in spherical harmonics:
\begin{equation}
 \phi=\sum_{\ell m}\frac{\Psi_\ell(r)}{\sqrt{r^2
+\tilde{a}^2 M^2}}e^{-i\omega t}Y^\ell(\vartheta,\varphi)\,,
\end{equation}
and expand the square root above to second order in $\tilde{a}$.
Schematically, we obtain the following equation:
\begin{equation}
  A_{\ell} Y^\ell+ D_\ell \cos^2\vartheta Y^\ell=0\,,\label{eq_expY}
\end{equation}
where a sum over $(\ell,m)$ is implicit, and the explicit form of the \textit{radial} coefficients $A_\ell$ and $D_\ell$ (which depend on $\phi$ and its radial derivatives) is given in the notebook~\cite{webpage,Pani:2012bp}. Note that the equation above can be seen as an expansion of Eq.~\eqref{KGtime} to ${\cal O}(\tilde a^2)$ and in the frequency domain.
The coefficient $D_\ell$ is proportional to $\tilde{a}^2$,
so the second term in the equation above is zero to first order in
rotation. Indeed, the angular dependence is already separated to first order, and the linearized Klein-Gordon equation can be cast in the form\footnote{\textbf{Exercise}: the formalism applies to a \emph{generic} stationary and axisymmetric background. Derive the field equation in the general background~\eqref{metric1}. To first order, it is easy to show that massive Klein-Gordon perturbations are described by:

\begin{equation}
 F B\Psi_{\ell}''+\frac{1}{2}\left[B'F+F'B\right]\Psi_{\ell}'
+\left[\omega^2-\frac{2 m \varpi(r) \omega}{r^2}-F\left(\frac{\ell(\ell+1)}{r^2}+\mu^2+
\frac{B'}{2r}+\frac{B F'}{2r F}\right)\right]\Psi_{\ell}=0\,.
\end{equation}
}
\begin{equation}
 \hat{\cal D}_2\Psi_\ell-\left[\frac{4m M^2
\tilde{a}\omega}{r^3}+F \frac{2M}{r^3}\right]\Psi_\ell=0\,,\label{scalar1st}
\end{equation}
where $F=1-2M/r$ and we defined the operator~\cite{Rosa:2011my}
\begin{equation}
 \hat{\cal D}_2=\frac{d^2}{d r_*^2}+\omega^2
-F\left[\frac{\ell(\ell+1)}{r^2}+\mu^2\right]\,,\label{D2}
\end{equation}
with $dr/dr_*=F$.
Equation~\eqref{scalar1st} coincides with Teukolsky's master
equation~\cite{Teukolsky:1973ha} for spin $s=0$ perturbations expanded
to first order in $\tilde{a}$.  

Let us separate the angular part of
Eq.~\eqref{eq_expY}. This can be achieved by using the
identity~\eqref{ident3} as well as the orthogonality properties of scalar spherical harmonics~\eqref{orthoscalar} [cf. also ~\ref{app:orthogonality}].
We obtain:
\begin{equation}
A_\ell+(\cQ_{\ell+1}^2+\cQ_{\ell}^2) D_\ell+\cQ_{\ell-1}\cQ_{\ell} D_{\ell-2}+\cQ_{\ell+2}\cQ_{\ell+1} D_{\ell+2 }=0\label{final_schematic}\,.
\end{equation}
Note that the coupling to perturbations with
indices $\ell\pm1$ is absent. This is due to the fact that Klein-Gordon
perturbations are polar quantities, and at first order the
Laporte-like selection rule implies that polar perturbations with
index $\ell$ should couple to axial perturbations with $\ell\pm1$, but
the latter are absent in the spin-0 case. At second order,
perturbations with harmonic index $\ell$ are coupled to perturbations
with the \emph{same} parity and $\ell\pm2$, but this coupling does not
contribute to the eigenfrequencies for the reasons discussed in the
previous section. We verify this statement below.

As discussed in Section~\ref{sec:time}, by repeated use of the identity~\eqref{ident1} we can separate the
perturbation equations at \emph{any order} in $\tilde{a}$. In the case of spin-1 or spin-2 perturbations, combinations of
vector and tensor spherical harmonics also appear, and these introduce terms such
as $(\sin\vartheta)^n Y^\ell_{,\vartheta}$. The latter, can be
decoupled in a similar fashion. For example, for computations to first and to second order, two useful relations are\footnote{\textbf{Exercise:} using the properties of the spherical harmonics, derive these identities.}
\begin{eqnarray}
\sin\th Y^{\ell}_{,\vartheta}&=&
{\cal Q}_{\ell+1}\ell Y^{\ell+1}-{\cal Q}_{\ell}(\ell+1)Y^{\ell-1}\,,\label{ident2}\\
\cos\th\sin\th  Y^{\ell}_{,\vartheta}&=&\left(\ell\cQ_\lp^2 -(\ell+1)\cQ_\ell^2\right)Y^\ell+\cQ_\lp \cQ_\lpp \ell Y^\lpp \nn\\
&&- \cQ_\ell \cQ_\lm (\ell+1) Y^\lmm\,.\label{ident4}
\end{eqnarray}
Using the explicit form of the coefficients given in the notebook, the field
equations~\eqref{final_schematic} schematically read
\begin{eqnarray}
\frac{d^2\Psi_\ell}{dr_*^2}+V_\ell \Psi_\ell+\tilde{a}^2&&
\Bigg[U_{\ell+2} \Psi_{\ell+2}+U_{\ell-2} \Psi_{\ell-2}+W_{\ell+2} \frac{d^2\Psi_{\ell+2}}{dr_*^2}+
W_{\ell-2} \frac{d^2\Psi_{\ell-2}}{dr_*^2}\Bigg]=0\,,\nn\\\label{final_tortoise}
\end{eqnarray}
where we have defined the tortoise coordinate via $dr/dr_*=
\Delta/(r^2+\tilde{a}^2M^2)$ (expanded at second order) and $V$, $U$ and $W$
are some potentials explicitly given in the notebook~\cite{webpage}.

As we expected, the coupling to the $\ell\pm2$ terms is proportional to
$\tilde{a}^2$. For a calculation accurate to second order in
$\tilde{a}$ the terms in parenthesis can be evaluated at zeroth order,
and the functions $\Psi_{\ell\pm2}^{(0)}$ must be solutions
of
\begin{equation}
 \frac{d^2\Psi_{\ell\pm2}^{(0)}}{dr_*^2}+V_{\ell\pm2}^{(0)} \Psi_{\ell\pm2}^{(0)}=0\,.
\end{equation}
By substituting these relations in Eq.~\eqref{final_tortoise} we get
\begin{eqnarray}
 \frac{d^2\Psi_\ell}{dr_*^2}+V_\ell \Psi_\ell&&+\tilde{a}^2
\left(U_{\ell+2}^{(0)}-V_{\ell+2}^{(0)}W_{\ell+2}^{(0)}\right) \Psi_{\ell+2}^{(0)}+\tilde{a}^2\left(U_{\ell-2}^{(0)}-
V_{\ell-2}^{(0)}W_{\ell-2}^{(0)}\right)  \Psi_{\ell-2}^{(0)}=0\,.\nn
\end{eqnarray}
Finally, making use of the expressions for $V$, $U$ and $W$, the field equations reduce to
\begin{eqnarray}
 &&\frac{d^2\Psi_\ell}{dr_*^2}+V_\ell \Psi_\ell=\frac{\tilde{a}^2 
M^2(r-2M)\left(\mu ^2-\omega ^2\right)}{r^3} \left[\cQ_{\ell+1} \cQ_{\ell+2} \Psi_{\ell+2}^{(0)}
+\cQ_{\ell-1} \cQ_{\ell} \Psi_{\ell-2}^{(0)}\right]\,,\nn\\\label{final}
\end{eqnarray}
where the potential is given by
\begin{eqnarray}
 V_\ell&&=\omega^2-F\left[
\frac{\ell(\ell+1)}{r^2}+\frac{2M}{r^3}+\mu^2\right]-\frac{4\tilde{a}m\omega M^2}{r^3}\nn\\
&&-\frac{\tilde{a}^2 M^2}{r^6}\left[24 M^2+4 M r 
\left(\ell(\ell+1)-3+r^2 \mu ^2\right)+r^4 F (\mu^2 -\omega^2 ) \left(\cQ_{\ell}^2
+\cQ_{\ell+1}^2\right)\right.\nn\\
&&\left.-2 M r^3 \omega ^2-r^2 \left(\ell(\ell+1)+m^2
+r^2 (\mu^2 -\omega^2 ) -1\right)\right]\,.\label{scalpot}
\end{eqnarray}
Note the similar structure of Eq.~\eqref{final} (which is valid to order $\tilde{a}^2$) and Eq.~\eqref{KGfinaltime} (which is valid to \emph{any} order in $\tilde{a})$.
Even if Eq.~\eqref{final} is approximate, one advantage of the slow-rotation approximation is that the couplings to terms with indices
$\ell\pm2$ can be neglected in the calculation of the modes to second order. In the
scalar case this can be shown explicitly as follows. If we define
\begin{equation}
 Z_\ell=\psi_\ell-\tilde{a}^2\left[c_{\ell+2}\psi_{\ell+2}-c_{\ell}\psi_{\ell-2}\right]\,, \label{Zell}
\end{equation}
where
\begin{equation}
c_{\ell}=\frac{M^2\left(\mu ^2-\omega ^2\right) \cQ_\lm \cQ_\ell}{2 (2\ell-1) }\,,
\end{equation}
then, at second order in rotation, Eq.~\eqref{final} can be written as a single, Schroedinger-like equation for $Z_\ell$:
\begin{equation}
 \frac{d^2 Z_\ell}{dr_*^2}+V_\ell Z_\ell=0\,,\label{final2}
\end{equation}
which can be solved by the methods discussed in Section~\ref{sec:time}. This equation
coincides with Teukolsky's master equation~\cite{Teukolsky:1973ha} for
spin $s=0$ perturbations expanded at second order in $\tilde{a}$. In
particular, the coefficients ${\cal Q}_{\ell}$ in Eq.~\eqref{scalpot}
agree with an expansion of Teukolsky's spheroidal eigenvalues to
second order in the BH spin~\cite{Berti:2005gp}\footnote{\textbf{Exercise:} by extending
our procedure to higher order, reconstruct the
Teukolsky scalar potential order by order in $\tilde{a}$.}. 

\subsection{Gravitational perturbations of a Kerr BH in the slow-rotation limit}\label{sec:Kerr}
As another relevant application of the slow-rotation expansion, here we derive the gravitational perturbations of a Kerr BH to first order in the angular momentum. Also in this case the linearized equations are known exactly in the Teukolsky formalism. Nonetheless, this exercise is propaedeutic to more involved cases that cannot be treated in the Teukolsky formalism. The entire procedure discussed in this section is presented in the notebook \url{slow_rot_grav_Kerr.nb}\cite{webpage}.

As a background, we consider the Kerr metric to first order in $\tilde{a}$, which is given in Eq.~\eqref{Kerr1st}. On this background we consider a harmonic decomposition of the metric perturbations as in Eqs.~\eqref{oddpart} and \eqref{evenpart} with $\eta_i^\ell\equiv G^\ell\equiv h_2^\ell\equiv0$.

Using this decomposition, we can solve Einstein's equations at linear order in the perturbations and to first order in $\tilde a$. Because of the transformation properties of the perturbation functions, the linearized Einstein equations naturally separate into three groups~\cite{Kojima:1992ie}. 
By denoting the linearized Einstein equations as $\delta {\cal E}_{\mu\nu}=0$, the first group reads
\begin{equation}
\delta{\cal E}_{(I)}\equiv (A^{(I)}_{{\ell}}+{\tilde A}^{(I)}_{{\ell}}\cos\th)Y^{{\ell}}
+B^{(I)}_{{\ell}}\sin\th Y^{{\ell}}_{,\vartheta}+C^{(I)}_{{\ell}} Y^{{\ell}}_{,\varphi}=0,\label{eqG1}
\end{equation}
where $I=0,1,2,3$ corresponds to $\delta {\cal E}_{tt}=0$, $\delta
{\cal E}_{tr}=0$, $\delta {\cal E}_{rr}=0$ and $\delta {\cal
  E}_{\vartheta\vartheta} +{\delta {\cal
    E}_{\varphi\varphi}}/{\sin\vartheta^2}=0$, respectively.
The second group reads
\begin{eqnarray}
&\delta{\cal E}_{(L\vartheta)}&\equiv(\alpha^{(L)}_{{\ell}}+{\tilde \alpha}^{(L)}_{{\ell}}\cos\th) Y^{{\ell}}_{,\vartheta}-
(\beta^{(L)}_{{\ell}}+{\tilde \beta}^{(L)}_{{\ell}}\cos\th)\frac{ Y^{{\ell}}_{,\varphi}}{\sin\th}\nn\\
&&+\eta^{(L)}_{{\ell}}\sin\th Y^{{\ell}}+\xi^{(L)}_{{\ell}}X^{{\ell}}+
\chi^{(L)}_{{\ell}}\sin\th W^{{\ell}}=0,\label{eqG2a}\\
&\delta{\cal E}_{(L\varphi)}&\equiv(\beta^{(L)}_{{\ell}}+{\tilde \beta}^{(L)}_{{\ell}}\cos\th) Y^{{\ell}}_{,\vartheta}+
(\alpha^{(L)}_{{\ell}}+{\tilde \alpha}^{(L)}_{{\ell}}\cos\th)\frac{ Y^{{\ell}}_{,\varphi}}{\sin\th}\nn\\
&&+\zeta^{(L)}_{{\ell}}\sin\th Y^{{\ell}}+\chi^{(L)}_{{\ell}}X^{{\ell}}-
\xi^{(L)}_{{\ell}}\sin\th W^{{\ell}}=0,\label{eqG2b}
\end{eqnarray}
where $L=0,1$ and the first equation corresponds to $\delta {\cal
  E}_{t\vartheta}=0$ and $\delta {\cal E}_{r\vartheta}=0$, whereas the
last equation corresponds to $\delta {\cal E}_{t\varphi}=0$ and
$\delta {\cal E}_{r\varphi}=0$.
Finally the third group is
\begin{eqnarray}
\delta {\cal E}_{(\vartheta\varphi)}&\equiv&	
f_{{\ell}}\sin\th Y^{{\ell}}_{,\th}+g_{{\ell}} Y^{{\ell}}_{,\varphi}+s_{{\ell}}
\frac{X^{{\ell}}}{\sin\th}+t_{{\ell}}W^{{\ell}}=0, \label{eqG3a}\\
\delta {\cal E}_{(-)}&\equiv&
g_{{\ell}}\sin\th Y^{\ell}_{,\th}-f_{{\ell}} Y^{{\ell}}_{,\varphi}-t_{{\ell}}
\frac{X^{{\ell}}}{\sin\th}+s_{{\ell}}W^{{\ell}}=0\,, \label{eqG3b}
\end{eqnarray}
corresponding to $\delta {\cal E}_{\vartheta\varphi}=0$ and $\delta {\cal  E}_{\vartheta\vartheta} -{\delta {\cal E}_{\varphi\varphi}}/{\sin\vartheta^2}=0$, respectively. In the equations above, $X^\ell$ and $W^\ell$ are the the tensor spherical harmonics defined as in Eq.~\eqref{XW}.

The coefficients appearing in these equations are \emph{linear} and \emph{purely radial} functions of the perturbation variables. Furthermore, they naturally divide into two sets accordingly to their parity:
 \begin{eqnarray}
 \text{Polar:}\qquad && A^{(I)}_{{\ell}},\quad C^{(I)}_{{\ell}},\quad \alpha^{(L)}_{{\ell}},\quad \tilde\beta^{(L)}_{{\ell}},\quad \zeta^{(L)}_{{\ell}},\quad \xi^{(L)}_{{\ell}},\quad s_{{\ell}},\quad f_{{\ell}},\nn\\
 \text{Axial:}\qquad &&\tilde A^{(I)}_{{\ell}},\quad B^{(I)}_{{\ell}},\quad \beta^{(L)}_{{\ell}},\quad \tilde\alpha^{(L)}_{{\ell}},\quad \eta^{(L)}_{{\ell}},\quad \chi^{(L)}_{{\ell}},\quad t_{{\ell}},\quad g_{{\ell}}.\nn
\end{eqnarray}
The explicit form of the
coefficients is given in the online notebook~\cite{webpage}. The crucial point is to recognize that the coefficients above are purely radial functions, that is, the entire angular dependence has been completely factored out in the linearized Einstein equations. 

\subsubsection{Separation of the angular dependence}
The decoupling of the angular dependence of the Einstein equations for
a slowly-rotating star was performed by Kojima~\cite{Kojima:1992ie} (see also~\cite{ChandraFerrari91}) by using
the orthogonality properties of the spherical harmonics, which are derived in \ref{app:orthogonality} for completeness. The procedure has been extended to the case of slowly-rotating BHs in Refs.~\cite{Pani:2012vp,Pani:2012bp}. Here we
adopt the same technique. 

Multiplying Eq.~\eqref{eqG1} by $Y^{*\,\ell'}$ and integrating over the sphere, we get
\begin{equation}
 0=\int d\Omega {Y^*}^\ell \delta{\cal E}_{(I)}=A_\ell^{(I)}+i m C_\ell^{(I)}+{\cal L}_0^{\pm1} \tilde A_{\ell}^{(I)}+{\cal L}_1^{\pm1} B_{\ell}^{(I)}\,,
\end{equation}
where the operators ${\cal L}_i^{\pm1}$ are defined in~\ref{app:orthogonality}. Using the explicit forms in the appendix, we obtain\footnote{\textbf{Exercise:} derive the radial equations \eqref{decG1},\eqref{decG2a}-\eqref{decG2b} and \eqref{decG3a}-\eqref{decG3b} explicitly.}
\begin{eqnarray}
&&A^{(I)}_{{\ell}}+i mC^{(I)}_{{\ell}}+\cQ_{{\ell}}\left[{\tilde A}^{(I)}_{{\ell-1}}
+(\ell-1){B}^{(I)}_{{\ell-1}}\right]+\cQ_{{\ell+1}}\left[{\tilde A}^{(I)}_{{\ell+1}}
-(\ell+2){ B}^{(I)}_{{\ell+1}}\right]=0,\nn\\\label{decG1}
\end{eqnarray}
where ${\cal Q}_{{\ell}}$ is defined as in Eq.~\eqref{Qpm}.
Similarly, Eqs.~\eqref{eqG2a}-\eqref{eqG2b} can be decoupled as follows
\begin{eqnarray}
 0&=&\int d\Omega\left[ {Y^*}^{\ell'}_{,\vartheta}\delta{\cal E}_{(L\vartheta)}+\frac{{Y^*}^{\ell'}_{,\varphi}}{\sin\vartheta}\delta{\cal E}_{(L\varphi)}\right]\\
&=& {\ell(\ell+1)}\alpha_{\ell}^{(L)}+{i} m\left[(\ell-1)(\ell+2)\xi^{(L)}_{{\ell}}
-{\tilde\beta}^{(L)}_{{\ell}}-\zeta^{(L)}_{{\ell}}\right]+{\cal L}_2^{\pm1}\eta_{\ell}^{(L)}+{\cal L}_3^{\pm1}\tilde\alpha_{\ell}^{(L)}+{\cal L}_4^{\pm1}\chi_{\ell}^{(L)}\,,\nn\\
 0&=&\int d\Omega\left[ {Y^*}^{\ell'}_{,\vartheta}\delta{\cal E}_{(L\varphi)}-\frac{{Y^*}^{\ell'}_{,\varphi}}{\sin\vartheta}\delta{\cal E}_{(L\vartheta)}\right]\\
&=&{\ell(\ell+1)} \beta^{(L)}_{{\ell}}+{i} m\left[(\ell-1)(\ell+2)\chi^{(L)}_{{\ell}}
+{\tilde\alpha}^{(L)}_{{\ell}}+\eta^{(L)}_{{\ell}}\right]+{\cal L}_2^{\pm1}\zeta_{\ell}^{(L)}+{\cal L}_3^{\pm1}\tilde\beta_{\ell}^{(L)}-{\cal L}_4^{\pm1}\xi_{\ell}^{(L)}\,.\nn
\end{eqnarray}
Using the explicit form of the operators ${\cal L}_i^{\pm1}$ given in~\ref{app:orthogonality}, the equations above reduce to
\begin{eqnarray}
 &&\ell(\ell+1) \alpha^{(L)}_{{\ell}}+i m\left[(\ell-1)(\ell+2)\xi^{(L)}_{{\ell}}
-{\tilde\beta}^{(L)}_{{\ell}}-\zeta^{(L)}_{{\ell}}\right]+\nn\\
&&\cQ_{{\ell}}(\ell+1)\left[(\ell-2)(\ell-1)\chi^{(L)}_{{\ell-1}}
+(\ell-1){\tilde\alpha}^{(L)}_{{\ell-1}}-\eta^{(L)}_{{\ell-1}}\right]-\nn\\
&&\cQ_{{\ell+1}}\ell\left[(\ell+2)(\ell+3)\chi^{(L)}_{{\ell+1}}
-(\ell+2){\tilde\alpha}^{(L)}_{{\ell+1}}-\eta^{(L)}_{{\ell+1}}\right]=0,
\label{decG2a}\\
&&\ell(\ell+1) \beta^{(L)}_{{\ell}}+i m\left[(\ell-1)(\ell+2)\chi^{(L)}_{{\ell}}
+{\tilde\alpha}^{(L)}_{{\ell}}+\eta^{(L)}_{{\ell}}\right]-\nn\\
&&\cQ_{{\ell}}(\ell+1)\left[(\ell-2)(\ell-1)\xi^{(L)}_{{\ell-1}}-
(\ell-1){\tilde\beta}^{(L)}_{{\ell-1}}+\zeta^{(L)}_{{\ell-1}}\right]+\nn\\
&&\cQ_{{\ell+1}}\ell\left[(\ell+2)(\ell+3)\xi^{(L)}_{{\ell+1}}
+(\ell+2){\tilde\beta}^{(L)}_{{\ell+1}}+\zeta^{(L)}_{{\ell+1}}\right]=0\,.
\label{decG2b}
\end{eqnarray}
%
Notice that Eqs.~\eqref{eqG1}-\eqref{eqG2b} have exactly the same form as Eqs.~(14)-(16) as in Kojima's original work~\cite{Kojima:1992ie}. Therefore, the radial equations have also the same schematic form. 
Finally, the last two equations~\eqref{eqG3a} and~\eqref{eqG3b} have the same form as Eqs.~(18)-(19) in Kojima's work~\cite{Kojima:1992ie} and their angular dependence can be eliminate by constructing the following relations
\begin{eqnarray}
 0&=&\int d\Omega \frac{1}{{\ell(\ell+1)}-2}\left[{W^*}^{\ell'}\delta{\cal E}_{(-)}+\frac{{X^*}^{\ell'}}{\sin\vartheta}\delta{\cal E}_{(\vartheta\varphi)}\right]={\ell(\ell+1)} s_{\ell}-im f_{\ell}+{\cal L}_2^{\pm1} g_{\ell} \,,   \nn\\
 0&=&\int d\Omega \frac{1}{{\ell(\ell+1)}-2}\left[{W^*}^{\ell'}\delta{\cal E}_{(\vartheta\varphi)}-\frac{{X^*}^{\ell'}}{\sin\vartheta}\delta{\cal E}_{(-)}\right]={\ell(\ell+1)} t_{\ell}+im g_{\ell}+{\cal L}_2^{\pm1} f_{\ell} \,,    \nn
\end{eqnarray}
which reduce to 
\begin{eqnarray}
&&0=\ell(\ell+1) s_{{\ell}}-i m f_{{\ell}}-\cQ_{{\ell}}(\ell+1)g_{{\ell-1}}+\cQ_{{\ell+1}}\ell g_{{\ell+1}} \label{decG3a}\,,\\
&&0=\ell(\ell+1) t_{{\ell}}+i m g_{{\ell}}-\cQ_{{\ell}}(\ell+1)f_{{\ell-1}}+\cQ_{{\ell+1}}\ell f_{{\ell+1}}\,. \label{decG3b}
\end{eqnarray}

To summarize, our decoupling procedure in the slow-rotation limit allows to obtain a system of $10$ coupled, ordinary differential equations. 
To first order, the mixing of different angular functions in Eqs.~\eqref{eqG1}-\eqref{eqG3b} results in a mixing of perturbation functions with multipolar indices $\ell$, $\ell+1$ and $\ell-1$ in the linearized radial equations.
The final radial equations are~\eqref{decG1},\eqref{decG2a}-\eqref{decG2b} and \eqref{decG3a}-\eqref{decG3b}. Their explicit form is available online~\cite{webpage}. 
Note that these equations have the general structure given in Eqs.~\eqref{epF1c}--\eqref{epF2c} to first order.
Clearly, not all ten linearized Einstein equations are independent and the coupled system can be simplified further.

\subsubsection{First-order corrections to the eigenvalue equations}
As previously discussed, the couplings to the $\ell\pm1$ terms
do not contribute to the QNM spectrum at first order in $\tilde a$. For
this reason we shall neglect these terms in the following. This allows us to treat the axial and polar sector separately.

Let us start with the axial sector. Neglecting the couplings to $\ell\pm1$ terms, the axial sector is fully described by three equations
\begin{eqnarray}
0&=&\ell(\ell+1) \beta^{(L)}_{{\ell}}+{i} m\left[(\ell-1)(\ell+2)\chi^{(L)}_{{\ell}}
+{\tilde\alpha}^{(L)}_{{\ell}}+\eta^{(L)}_{{\ell}}\right] ,\nn\\ \label{axial_dec}
0&=&\ell(\ell+1) t_{{\ell}}+{i} m g_{{\ell}} , \label{dec_t}
\end{eqnarray}
where $L=0,1$. Only two equations are independent and they can be solved for the axial perturbations $h_0^\ell$ and $h_1^\ell$. 
We define the Regge-Wheeler function $\Psi^\ell$ as
\begin{equation}
 h_1^\ell=\frac{r\Psi^\ell}{F},
\end{equation}
Then, from Eq.~\eqref{axial_dec} with $L=1$, we get
\begin{eqnarray}
 {h_0^\ell}'&=&\frac{2 (r-2M) \omega  h_{00}^\ell-i \left(\left(\ell(\ell+1)-2\right) (2 M-r)+r^3 \omega ^2\right) \Psi^\ell}{r (r-2M) \omega }\nn\\
&&+\frac{2 \tilde{a}M^2 m \left(6 (r-2M) \omega  h_{00}^\ell-i \ell (\ell+1) \left(\left(\ell(\ell+1)-2\right) (2 M-r)-r^3 \omega ^2\right) \Psi^\ell\right)}{\ell(\ell+1) r^4 (r-2M) \omega ^2}\,.\nn
\end{eqnarray}
Substituting this equation in the remaining Eqs.~\eqref{axial_dec} with
$L=0$ and $L=2$ we get a decoupled equation for $\Psi^\ell$ only. By defining
\begin{equation}
 \Psi^\ell=\psi^\ell\left(1-\frac{2m \tilde{a}M^2}{\omega r^3}\right)\,,
\end{equation}
the final modified Regge-Wheeler equation describing axial perturbations reads
\begin{equation}
 \frac{d^2\psi^\ell}{dr_*^2}+\left[\omega^2-\frac{4m \tilde{a}M^2\omega}{r^3}-V_{\rm axial}\right]\psi^\ell=0\,.\label{RWaxial}
\end{equation}
with
\begin{equation}
 V_{\rm axial}=F(r)\left(\frac{\ell(\ell+1)}{r^2}-\frac{6M}{r^3}+\frac{24m\tilde{a}M^2(3r-7M)}{\ell(\ell+1)\omega r^6}\right)\,.
\end{equation}
The Schroedinger-like equation above can then be solved with standard methods, e.g. continued fractions or direct integration.

Let us now turn to the polar sector, which is slightly more involved. Neglecting the coupling to axial perturbations with $\ell\pm1$, the polar sector is described by the following equations:

\begin{eqnarray}
0&=&A^{(I)}_{{\ell}}+i mC^{(I)}_{{\ell}}\,,\\
0&=&\ell(\ell+1) \alpha^{(L)}_{{\ell}}+i m\left[(\ell-1)(\ell+2)\xi^{(L)}_{{\ell}}
-{\tilde\beta}^{(L)}_{{\ell}}-\zeta^{(L)}_{{\ell}}\right]\,,\\
0&=&\ell(\ell+1) s_{{\ell}}-i m f_{{\ell}} \,.
\end{eqnarray}
The last equation can be solved for $H_2^\ell$, whereas the first equation with $I=2$ can be solved for $H_0^\ell$. The remaining equations can be reduced to a system of first order equations for $K^\ell$ and $H_1^\ell$. Finally, the system can be reduced to a single second-order equation for a new function $Z^\ell$ in term of which $K^\ell$ and $H_1^\ell$ are defined. The detailed procedure is derived in the notebook~\url{slow_rot_grav_Kerr.nb}~\cite{webpage}. The final result is a modified Zerilli equation describing polar perturbations:
\begin{equation}
 \frac{d^2Z^\ell}{dr_*^2}+\left[\omega^2-\frac{4m \tilde{a}M^2\omega}{r^3}-V_{\rm polar}\right]Z^\ell=0\,.\label{Zerillipolar}
\end{equation}
with
\begin{eqnarray}
 V_{\rm polar}&&=F(r)\left[\frac{2 M}{r^3}+\frac{(\ell-1) (\ell+2)}{3} \left(\frac{1}{r^2}+\frac{2 (\ell-1) (\ell+2) \left(\ell(\ell+1)+1\right)}{\left(6 M+r \left(\ell(\ell+1)-2\right)\right)^2}\right)\right.\nn\\
 &&\left.+\frac{4 \tilde{a} m M^2 }{r^7 \ell  (\ell+1 ) \left(6 M+r \left(\ell(\ell+1)-2\right)\right)^4 \omega } \left(27648 M^6+2592 M^5 r (6 \ell  (\ell+1 )-19)\right.\right.\nn\\
 &&\left.\left.+144 M^4 r^2 \left(230+\ell  (\ell+1 ) (21 \ell  (\ell+1 )-148)+6 r^2 \omega ^2\right)\right.\right.\nn\\
 &&\left.\left.+12 M^2 r^4 \left(\ell(\ell+1)-2\right)^2\left(\ell  (\ell+1 ) (-12+5 \ell  (\ell+1 ))+28 r^2 \omega ^2-4\right)\right.\right.\nn\\
 &&\left.\left.+12 M^3 r^3 \left(\ell(\ell+1)-2\right) \left(374+\ell  (\ell+1 ) (29 \ell  (\ell+1 )-200)+72 r^2 \omega ^2\right)\right.\right.\nn\\
 &&\left.\left.+r^6 \left(\ell(\ell+1)-2\right)^3 \left(-3 \left(\ell(\ell+1)-2\right) \left(\ell(\ell+1)+2\right)+2 r^2 \left(\ell(\ell+1)-4\right) \omega ^2\right)\right.\right.\nn\\
 &&\left.\left.+M r^5 \left(\ell(\ell+1)-2\right)^2 \left(\left(\ell(\ell+1)-2\right) \left(\ell(\ell+1)+2\right) (7 \ell  (\ell+1 )-38)\right.\right.\right.\nn\\
 &&\left.\left.\left.+24 r^2 (2 \ell  (\ell+1 )-5) \omega ^2\right)\right)\right]\,.
\end{eqnarray}
Although the polar potential is more involved than the axial one, we are still left with a single second-order ODE that can be solved by standard methods\footnote{\textbf{Exercise:} compute the polar and axial QNMs integrating Eqs.~\eqref{RWaxial} and \eqref{Zerillipolar}. The axial eigenvalue problem can be reduced to a six-term recurrence relation and solved by continued fractions. The polar sector can be solved by direct integration. Interestingly, the QNM spectrum is the same for the two sectors. This extends to first order in the rotation the well-known fact that axial and polar perturbations of a Schwarzschild BH are isospectral~\cite{Chandra} and it is consistent with a mode analysis of the Kerr metric in the Teukolsky formalism.}.

\subsection{Computing the QNMs in slow-rotation limit}
One of the key advantages of the slow-rotation approximation is that, for basically any stationary and axisymmetric background, the
linearized field equations have a form which is very similar to the nonrotating case. Thus, all existing methods to solve the linear dynamics around spherically symmetric BHs can be directly applied to the slow-rotation case with only minor modifications. The two principal modifications are:
\begin{itemize}
 \item[i)] The coupling between modes with different parity and differnt harmonic indices $\ell$,
 \item[ii)] The behavior of the fields close to the horizon.
\end{itemize}
The first correction automatically implies that, in the Fourier space, we are dealing with coupled systems of ODEs in the form:
\begin{equation}
 {\cal D}\mathbf{Y}^\ell +\mathbf{V}\mathbf{Y}^\ell=\mathbf{S}_{+1}\mathbf{Y}^{\ell+1}+\mathbf{S}_{-1}\mathbf{Y}^{\ell-1}+\mathbf{S}_{+2}\mathbf{Y}^{\ell+2}+\mathbf{S}_{-2}\mathbf{Y}^{\ell-2}+\dots\,. \label{systemC}
\end{equation}
where ${\cal D}$ is some differential radial operator, $\mathbf{V}$ and $\mathbf{S}_{\pm n}$ are radial $N\times N$ matrices and $\mathbf{Y}^{\ell}$ is a $N$-dimensional vector which contains all relevant perturbation functions with harmonic index $\ell$.

Since $\ell=0,1,2,..$, the full system~\eqref{systemC} formally contains an \emph{infinite} number of equations.
In practice, we can truncate it at some given value of $\ell=L$, compute the modes
as explained below, and finally check convergence by increasing the truncation
order. 

Let us suppose we truncate the couplings at order $\ell+p$, i.e. for a given $m$ we assume
\begin{equation}
 \mathbf{Y}^{L}\equiv0\qquad {\rm when}\qquad L>\ell+p\,.
\end{equation}
Therefore, we can recast the system~\eqref{systemC} into a system of $(2p+1)N$ coupled second-order ODEs or, equivalently, to a system of $(2p+1)2N$ first-order ODEs in the schematic form
\begin{equation}
 \frac{d \mathbf{Z}}{dr_*}+\mathbf{W}\mathbf{Z}=0\,, \label{system1st}
\end{equation}
where $r_*$ is some suitable coordinate and the $(2p+1)2N$-dimensional vector $\mathbf{Z}$ contains $\mathbf{Y}^{\ell}$, $\mathbf{Y}^{\ell\pm 1}$,..., $\mathbf{Y}^{\ell\pm p}$ and their first derivatives. A consistency check of the slow-rotation approximation is that the couplings to perturbations with $\ell-p$ are automatically vanishing when $p>\ell$. Although the system~\eqref{system1st} can contain several coupled equations, nonetheless the latter are ODEs and the corresponding eigenvalue problem (or the corresponding time evolution in the time-domain) can be analyzed with the methods discussed in the previous sections. 

The second modification listed above arises due to the frame-dragging effect. Indeed, it is generically possible to recast the perturbation functions and to choose a suitable radial coordinate $r_*$ such that
\begin{eqnarray}
 Y_i^\ell(r)&\sim& e^{i\omega r_*},\qquad \hspace{0.4cm}r\to\infty, \label{asym_inf}\\
 Y_i^\ell(r)&\sim& e^{-i k_H r_*},\qquad r\to r_+, \label{asymp_hor}
\end{eqnarray}
where $r$ is the original radial coordinate as defined in Eq.~\eqref{metric2},
\begin{equation}
k_H=\omega-m\Omega_H,\label{kH}
\end{equation}
and 
\begin{equation}
 \Omega_H=-\lim_{r\to r_+}\frac{g^{(0)}_{t\varphi}}{g^{(0)}_{\varphi\varphi}}\,,
\end{equation}
is the angular velocity at the horizon of locally non-rotating observers to some given order in $\tilde{a}$. 
The near-horizon behavior~\eqref{asymp_hor} shows that, if $\omega<m\Omega_H$, an observer at infinity would see waves outgoing from the horizon. 
By computing the energy and angular momentum fluxes carried by these waves, it is possible to show that superradiant amplification occurs in spinning BH spacetimes when the superradiance condition $k_H<0$ is met. This analysis was originally performed for perturbations of a Kerr BH within the Teukolsky formalism~\cite{Teukolsky:1974yv}. We have here shown how the same result can be obtained for any perturbation of generic rotating, axisymmetric BH spacetimes within the slow-rotation approximation. An important point to bear in mind is that, in order to consistently discuss superradiance in a slow-rotation approximation, one needs to include at least \emph{second-order} terms in the expansion~\cite{Pani:2012bp}. Indeed, at superradiance $\omega<\Omega_H\sim{\cal O}(\tilde a)$ and the energy of the wave scales as $\omega^2\sim{\cal O}(\tilde a^2)$, so that a first-order analysis is in principle not sufficient. Nonetheless, at least in some specific case~\cite{Pani:2012bp}, the first and second order results are in qualitatively (and sometimes in remarkably good quantitative) agreement even in the superradiant regime where, in principle, deviations of order unity might be expected. One naive reason for this agreement is that, by symmetry arguments, the superradiant condition itself only contains odd powers of $\tilde{a}$, so that the first-order corrections is valid up to third order and the onset of superradiance can be captured already at first order.

\subsubsection{Eigenvalue problems with couplings to different $\ell$: Breit-Wigner method}
When considering the couplings to higher multipolar indices, the number of coupled equations that have to be solved may be quite large, depending on the order of the slow-rotation expansion and on the number of perturbation variables for a given $\ell$. In principle, any of the methods previously discussed for coupled systems can be applied to this case. In practice, if the number of equations is large, some of the methods become inefficient. 

If the eigenvalue problem admits slowly-damped modes (i.e. modes with $\omega_I\ll\omega_R$) then the Breit-Wigner method discussed above is very convenient because its simplicity makes it well-suited to deal with large systems of ODEs.

Roughly speaking, slowly-damped modes exist if the potential has a sufficiently deep minimum. In a single-ODE problem this happens, for example, if the perturbation field is massive~\cite{Detweiler:1980uk,Dolan:2007mj} or for small AdS BHs~\cite{Berti:2009wx}. In a coupled-ODE problem the situation is less clear, but slowly-rotating modes are expected in the same settings and indeed they have been found recently~\cite{Rosa:2011my,Pani:2012vp,Pani:2012bp}.

To illustrate how the matrix-valued Breit-Wigner method works in case of couplings with different harmonic indices $\ell$, let us consider one simple case: massive spin-1 (Proca) perturbations of a Kerr BH to second order in the rotation~\cite{Pani:2012bp}. The method has been implemented in the notebook~\url{BW_Proca_2nd_order.nb}~\cite{webpage}.

The coupled system contains three ODEs for two polar functions $u_{(2)}^\ell$ and $u_{(3)}^\ell$ and an axial function $u_{(4)}^\ell$. As explained before, at first order in $\tilde{a}$ the polar functions are coupled to $u_{(4)}^{\ell\pm1}$ and, at second order in $\tilde{a}$, they are also coupled to $u_{(2)}^{\ell\pm2}$ and $u_{(3)}^{\ell\pm2}$.
Let us suppose we truncate the axial sector at $\ell=L$ and the polar
sector at $\ell=L+1$. When $m=0$, the system reduces to $N=3L$ coupled
second-order ODEs for $L-1$ axial functions and $2L+1$ polar
functions, including the monopole~\cite{Pani:2012bp}. When $|m|>0$ the truncated system contains
$N=3L-3|m|+2$ second-order ODEs (for $L-|m|$ axial functions and
$2L-2|m|+2$ polar functions). In all cases we are left with a system
of $N$ second-order ODEs for $N$ perturbation functions\footnote{\textbf{Exercise:} as we have discussed, to second order in $\tilde{a}$, only the couplings to $\ell\pm1$ are important. This property can be verified numerically by truncating the coupled system to some high order and checking that the modes do not change if the truncation order is greater than $\ell+1$.}.

In this case the Breit-Wigner method proves to be very instructive, because it makes manifest an interesting physical property of the Proca system~\cite{Pani:2012bp}.
When the mass of the spin-1 field is small, $M\mu\ll1$, the spectrum has a hydrogen-like behavior:
\begin{equation}
 \omega_R^2=\mu^2\left[1-\left(\frac{M\mu}{\ell+n+S +1}\right)^2\right]+{\cal O}\left(\mu^4\right)\,,\label{wRProca}
\end{equation}
where $n\geq0$ is the overtone number and $S$ is the so-called polarization index [in the Proca case, $S=0$ for transverse axial modes and $S=\pm1$ for longitudinal polar modes]. The equation above predicts an approximate degeneracy for modes with the same value of
$\ell+n+S$ in the small $\mu$ limit. In the Breit-Wigner method, the mode frequencies
can be identified as minima of the real-valued function
$|\det{\mathbf{S}}|^2$.  The approximate degeneracy translates into a series of minima which are very close to each other in the real axis (in fact, their separation scales with $\mu^4$ in the small $\mu$ limit). 

In the notebook~\url{BW_Proca_2nd_order.nb} we show $|\det{\mathbf{S}}|^2$ in a given range of $\omega_R$ for $m=1$ and arbitrary truncation order~\cite{webpage}. In the selected range, the function shows three minima, which correspond to a three-fold degeneracy, $\ell+n+S=1$ in Eq.~\eqref{wRProca}. The latter can be achieved by $(\ell,n,S)=(1,0,0)$, $(2,0,-1)$, $(1,1,-1)$. Therefore, one minimum corresponds to the axial fundamental mode, whereas the other two minima correspond to the fundamental and to the first overtone of the polar modes. This example shows the advantage of the resonance method: in a single numerical implementation one is able to obtain the \emph{full} quasi-bound spectrum, irrespectively of the parity of the modes and even for different values of $\ell$, up to a given truncation order.

\section{Conclusions and open problems}	\label{sec:conclusions}
We have presented self-consistent tools to derive, separate and solve numerically the perturbation equations of stationary and axisymmetric BHs within some approximate scheme. 
In general, the linearized field equations on a stationary and axisymmetric spacetime form a coupled $2+1$ dimensional system which can be evolved using the techniques discussed elsewhere~\cite{Witek:2012tr,Helvi,Hiro}. Such evolution requires advanced numerical methods and it is usually time- and resource-consuming. Here, we have introduced complementary techniques to reduce the problem to a $1+1$ time evolution~\cite{Dolan:2012yt} or to a simple one dimensional problem in the frequency domain.

Working in some perturbative scheme, it is possible to solve the linear dynamics on \emph{generic} stationary and axisymmetric BHs. This requires a combination of analytical and numerical tools, which we have discussed in some detail. 

Most of the numerical methods presented in this work have been implemented in simple {\scshape Mathematica} notebooks which are publicly available~\cite{webpage}. We hope this will help students and researchers to adapt and extend them to a multitude of problems that are still open in this field. 

Spinning BHs play a crucial role in gravity and in astrophysics, and they are indeed ubiquitous in modern applications.
There are several venues in which the techniques we discussed can be useful.
The linear response of a BH to gravitational perturbations is mostly unknown if the background metric is not Kerr or if the underlying theory of gravity is not GR. These extensions are important to discuss the BH ringdown and the gravitational-wave emission in modified theories of gravity. 
On a more theoretical side, the linear dynamics of hairy BHs is relevant in the context of the gauge/gravity duality. If one wishes to describe an axisymmetric theory in the holographic space, understanding the linear dynamics of spinning AdS BHs is crucial.
Probably one of the most important applications concerns the study of BH perturbations in higher-dimensions. Besides some particular cases of enhanced symmetries, a generic treatment of perturbed spinning BHs in higher dimensions is still lacking. This prevents a complete stability analysis and a full understanding of the greybody factors of spinning higher-dimensional BHs.
Here we have focused on vacuum solutions, but astrophysical BHs are surrounded by various type of matter and magnetic fields. Extending the techniques discussed in this work to the case of nonvacuum solutions is an interesting problem.
Finally, some direct applications of the slow-rotation approach include the study of massive spin-2 fields around spinning BHs~\cite{Brito:2013wya} and the analysis of gravito-electromagnetic perturbations of the Kerr-Newman metric in GR~\cite{Pani:2013ija,Pani:2013wsa}. We hope to report on these interesting topics in the near future.

\section*{Acknowledgments}
We thank Leonardo Gualtieri for suggestions and for a careful reading of the manuscript. 
It is also a pleasure to thank Emanuele Berti, Vitor Cardoso and Leonardo Gualtieri, who have been involved in various projects related to this work, and the Organizer and the Editors of the NR/HEP2 Spring School~\cite{webpage}.
This work was supported by the Intra-European Marie Curie contract aStronGR-2011-298297 and by FCT - Portugal through PTDC projects FIS/098025/2008, FIS/098032/2008, CERN/FP/123593/2011. 
Computations were performed on the ``Baltasar Sete-Sois'' cluster at IST,
the cane cluster in Poland through PRACE DECI-7 ``Black hole dynamics
in metric theories of gravity'',  
on Altamira in Cantabria through BSC grant AECT-2012-3-0012,
on Caesaraugusta in Zaragoza through BSC grants AECT-2012-2-0014 and AECT-2012-3-0011,
XSEDE clusters SDSC Trestles and NICS Kraken
through NSF Grant~No.~PHY-090003, Finis Terrae through Grant
CESGA-ICTS-234.

\appendix
\section{List of publicly available codes~\cite{webpage}}
Most of the numerical and analytical methods discussed in the main text have been directly implemented in ready-to-be-used  {\scshape Mathematica}\textsuperscript{\textregistered} notebooks, which are publicly available~\cite{webpage}.
Here we give a short description of them:
\begin{itemize}
 \item \url{BW_Proca_2nd_order.nb}: Proca quasi-bound states of a Kerr BHs in GR to second order in the spin (including couplings with different harmonic indices) via the Breit-Wigner resonance method.
 \item \url{CF_matrix_3terms.nb}: scalar, electromagnetic and gravitational QNMs of a Schwarzschild BH computed with a matrix-valued continued fraction method. 
 \item \url{DCS_DI.nb}: Gravito-scalar QNMs of Schwarzschild BHs in Dynamical Chern-Simons gravity computed with a matrix-valued direct integration.
 \item \url{DCS_pert_eqs.nb}: Derivation of the gravito-scalar perturbation equations of Schwarzschild BHs in Dynamical Chern-Simons gravity.
 \item \url{field_eqs.nb}: Derivation of the field equations starting from a Lagrangian in a modified gravity theory.
 \item \url{series_method_DCS.nb}: Gravito-scalar QNMs of Schwarzschild-AdS BH in Dynamical Chern-Simons gravity computed with a matrix-valued series method.
 \item \url{slow_rot_grav_Kerr.nb}:  Derivation of the gravitational perturbation equations (axial and polar) of a Kerr background to first order in the angular momentum.
 \item \url{slow_rot_scalar.nb}: Derivation of the perturbation equations for a massive Klein-Gordon equation on a Kerr background to second order in the angular momentum.
\end{itemize}
\section{Orthogonality properties of the spherical harmonics} \label{app:orthogonality}
In this appendix we give some useful orthogonality properties of scalar, vector and tensor spherical harmonics. For clarity, here we explicitly append both multipolar indices $\ell$ and $m$.
We define the scalar product on the two--sphere as
\begin{eqnarray}
<f,g>&\equiv&\int d\Omega f^*g=\int d\vartheta d\varphi\sin\vartheta f^*g\,, \label{scalarproduct}\\
<f_a,g_a>&\equiv&\int d\Omega f_a^*g_b\gamma^{ab}\,, \\
<f_{ab},g_{ab}>&\equiv&\int d\Omega f_{ab}^*g_{cd}\gamma^{ca}\gamma^{db}\,,
\end{eqnarray}
where $\gamma_{ab}={\rm diag}(1,\sin^2\vartheta)$ is the induced metric on the two-sphere.
By definition, scalar spherical harmonics satisfy the fundamental identity
\begin{equation}
Y^{\ell m}_{,\vartheta\vartheta}+\cot\vartheta Y^{\ell m}_{,\vartheta}+
\frac{1}{\sin^2\vartheta}Y^{\ell m}_{,\varphi\varphi}=-\ell(\ell+1) Y^{\ell m}\,. \label{propY}
\end{equation}
From this equation and from the orthogonality property for the scalar 
spherical harmonic [cf. also Eq.~\eqref{orthoscalar}]:
\begin{equation}
<Y^{\ell m},Y^{\ell'm'}>=\delta^{\ell\ell'}\delta^{mm'}\,,
\label{orthoY}
\end{equation}
we obtain the following relations:
\begin{eqnarray}
<Y^{\ell m}_a,Y^{\ell m}_a>&=&<S^{\ell m}_a,S^{\ell m}_a>=\int d\vartheta d\varphi\sin\vartheta\left(Y^{*\ell\,m}_{,\vartheta}Y^{\ell m}_{,\vartheta}
+\frac{1}{\sin^2\vartheta}Y^{*\ell\,m}_{,\varphi}Y^{\ell m}_{,\varphi}\right)\nn\\
&&=-\int d\vartheta d\varphi\sin\vartheta Y^{*\ell\,m}\left(Y^{\ell m}_{,\vartheta\vartheta}+
\cot\vartheta Y^{\ell m}_{,\vartheta}+\frac{1}{\sin^2\vartheta}Y^{\ell m}_{,\varphi\varphi}\right)\nn\\
&&=\ell(\ell+1) \int d\vartheta d\varphi\sin\vartheta Y^{*\ell\,m}Y^{\ell m}\nn\\
&&=2(n+1)\,,
\label{orthoYS}
\end{eqnarray}
where $2n\equiv(\ell-1)(\ell+2)$ and we have defined the polar and axial vector harmonics, which respectively read:
\begin{eqnarray}
 Y_{a}^{\ell m}&=&(Y^{\ell m}_{,\vartheta},Y^{\ell m}_{,\varphi})\,,\\
 S_{a}^{\ell m}&=&(- Y^{\ell m}_{,\varphi}/\sin\vartheta,\sin\vartheta Y^{\ell m}_{,\vartheta})\,.
\end{eqnarray}

%
The spin--two harmonics are defined as
\begin{equation}
_{-2}S^{\ell m}(\vartheta,\varphi)\equiv\frac{W^{\ell m}(\vartheta,\varphi)-{i}X^{\ell m}(\vartheta,\varphi)/\sin\vartheta}{\ell(\ell+1) (\ell(\ell+1)-2)}\,,
\end{equation}
where $W^{\ell m}$ and $X^{\ell m}$ are defined as in Eqs.~\eqref{XW}. The spin--two harmonics satisfy the orthogonality property:
\begin{equation}
<\,_{-2}S^{\ell m},\,_{-2}S^{\ell'm'}>=\delta^{\ell\ell'}\delta^{mm'}\,.
\end{equation}
Using this relation, one can obtain the following:
\begin{eqnarray}
&&\frac{1}{2}<Z^{\ell m}_{ab},Z^{\ell'm'}_{ab}>=
\frac{1}{2}<S^{\ell m}_{ab},S^{\ell'm'}_{ab}>\nn\\
&&=\int d\vartheta d\varphi\sin\vartheta\left(W^{*\ell\,m}_{ab}W^{\ell'm'}_{cd}
+\frac{X^{*\ell\,m}_{ab}X^{\ell'm'}_{cd}}{\sin^2\vartheta}\right)\gamma^{ca}\gamma^{db}\nn\\
&&=4n(n+1)\delta^{\ell\ell'}\delta^{mm'}\,,\label{orthoZS}
\end{eqnarray}
where we have defined the polar and axial tensor harmonics, which respectively read:
\begin{eqnarray}
 Z_{ab}^{\ell m}&=&\left(\begin{array}{cc}
                          W^{\ell m}&\quad  X^{\ell m}\\
                          X^{\ell m}&\quad  -\sin^2\vartheta W^{\ell m}
                         \end{array}\right)\,,\\
 S_{ab}^{\ell m}&=&\left(\begin{array}{cc}
                          -X^{\ell m}/\sin\vartheta&\quad \sin\vartheta W^{\ell m}\\
                          \sin\vartheta W^{\ell m}&\quad  \sin\vartheta X^{\ell m}
                         \end{array}\right)\,,\,.
\end{eqnarray}

Similarly, we have 
\begin{eqnarray}
\int d\Omega \left[{W^*}^{\ell' m'} Y^{\ell m}_{,\varphi}-{X^*}^{\ell'm'} Y^{\ell \,m}_{,\vartheta}\right]&=&i m(\ell(\ell+1)-2)\delta_{mm'}\delta_{\ell\ell'}\,,\nn\\
\int d\Omega \cos\vartheta\left[{W^*}^{\ell' m'}W^{\ell\,m}+\frac{{X^*}^{\ell'm'}X^{\ell\,m}}{\sin\vartheta^2}\right]&=&2im(\ell(\ell+1)-2)\delta_{mm'}\delta_{\ell\ell'}\,,\nn\\
\int d\Omega \left[\frac{{{W^*}^{\ell'm'}X^{\ell\,m}}-{{X^*}^{\ell'm'}W^{\ell\,m}}}{\sin\vartheta}\right]&=&0\,.\nn
\end{eqnarray}

Two other useful identities involving the spherical harmonics are given in Eqs.~\eqref{ident1} and \eqref{ident2}.
Using those identities, we can evaluate the following operators, acting on a generic function $A_{\ell m}$
\begin{eqnarray}
{\cal L}_0^{\pm1} A_{\ell m} &\equiv& A_{\ell'm'}\int d\Omega {Y^*}^{\ell m}\cos\vartheta Y^{\ell' m'}={\cal Q}_{\ell m} A_{\ell-1 m}+{\cal Q}_{\ell+1 m}A_{\ell+1 m}	\,,\nn\\
{\cal L}_1^{\pm1} A_{\ell m} &\equiv& A_{\ell'm'} \int d\Omega {Y^*}^{\ell m}\sin\vartheta  Y^{\ell' m'}_{,\vartheta}=(\ell-1){\cal Q}_{\ell m} A_{\ell-1 m}-(\ell+2){\cal Q}_{\ell+1 m} A_{\ell+1 m}		\,,\nn\\
{\cal L}_2^{\pm1} A_{\ell m} &\equiv& \left[-2{\cal L}_0^{\pm1}-{\cal L}_1^{\pm1}\right] A_{\ell m} =-(\ell+1){\cal Q}_{\ell m} A_{\ell-1 m}+\ell Q_{\ell+1 m} A_{\ell+1 m}		\,.\nn\\
{\cal L}_3^{\pm1} A_{\ell m} &\equiv& \left[{\ell(\ell+1)}{\cal L}_0^{\pm1}+{\cal L}_1^{\pm1}\right] A_{\ell m} \nn\\
&=&(\ell-1)(\ell+1){\cal Q}_{\ell m} A_{\ell-1 m}+\ell(\ell+2){\cal Q}_{\ell+1 m} A_{\ell+1 m}		\,,\nn\\
{\cal L}_4^{\pm1} A_{\ell m} &\equiv&\left[-2({\ell(\ell+1)}-2){\cal L}_0^{\pm1}+({\ell(\ell+1)}+2){\cal L}_1^{\pm1}\right]A_{\ell m} \nn\\
&=&(\ell^2-1)(\ell-2){\cal Q}_{\ell m} A_{\ell-1 m}-\ell(\ell+2)(\ell+3)Q_{\ell+1 m} A_{\ell+1 m}\,,\nn
\end{eqnarray}
where ${\cal Q}_{\ell m}$ is defined as in Eq.~\eqref{Qpm} (omitting the subscript $m$). The operators above are used in the main text to separate the angular dependence of the linearized field equations within the slow-rotation expansion.
\bibliographystyle{ws-ijmpa}
\bibliography{slowrot}
  
\end{document}